\documentclass[twocolumn, notitlepage, 10pt, aps, floatfix, showpacs, prb, citeautoscript, superscriptaddress]{revtex4-1}
\usepackage{graphicx}
\usepackage{amsmath, amssymb}
\usepackage{bm}
\usepackage{xcolor}
\usepackage[colorlinks, citecolor={blue!50!black}, urlcolor={blue!50!black}, linkcolor={red!50!black}]{hyperref}
\usepackage{bookmark}
\usepackage[version=4]{mhchem}
\usepackage[load=physical,load=abbr]{siunitx}
\usepackage[normalem]{ulem}
\usepackage{microtype}

\newcommand{\pmat}[1]{\begin{pmatrix}#1\end{pmatrix}}
\newcommand{\avg}[1]{\langle#1\rangle}
\newcommand{\comment}[1]{}

\begin{document}
\begin{abstract}
Semiconductors in the proximity of superconductors have been proposed to support phases hosting Majorana bound states.
When the systems undergo a topological phase transition towards the Majorana phase, the spectral gap closes, then reopens, and the quasiparticle band spin polarization is inverted.
We focus on two paradigmatic semiconductor-superconductor heterostructures and propose an all-electrical spectroscopic probe sensitive to the spin inversion at the topological transition. 
Our proposal relies on the indirect coupling of a time-dependent electric field to the electronic spin due to the strong Rashba spin-orbit coupling in the semiconductor.
We analyze within linear response theory the dynamical correlation functions and demonstrate that some components of the susceptibility can be used to detect the nontrivial topological phases.
\end{abstract}

\title{All-electrical spectroscopy of topological phases in semiconductor-superconductor heterostructures}
\author{Doru Sticlet}
\email{corresp: doru.sticlet@itim-cj.ro}
\affiliation{National Institute for Research and Development of Isotopic and Molecular Technologies, 67-103 Donat, 400293 Cluj-Napoca, Romania}
\author{C\u at\u alin Pa\c scu Moca}
\affiliation{MTA-BME Quantum Dynamics and Correlations Research Group, Institute of Physics, Budapest University of Technology and Economics, Budafoki ut 8., H-1111 Budapest, Hungary}
\affiliation{Department  of  Physics,  University  of  Oradea,  410087,  Oradea,  Romania}
\author{Bal\'azs D\'ora}
\affiliation{MTA-BME Lend\"ulet Topology and Correlation Research Group, Budapest University of Technology and Economics, 1521 Budapest, Hungary}
\affiliation{Department of Theoretical Physics, Budapest University of Technology and Economics, 1521 Budapest, Hungary}
\maketitle

\section{Introduction}
\label{sec:intro}
\comment{Majorana is a hot topic}
There has been a growing interest in the condensed matter scientific community in exploring topological superconducting phases supporting Majorana bound states, partially motivated by the prospects of a quantum computer.
Majorana bound states are quasiparticles in condensed matter theory which are their own antiparticle and possess non-Abelian statistics.
They may appear unpaired as zero-energy excitations, energetically separated from the quasiparticle continuum by a superconducting gap~\cite{Kitaev2001}.
Within the restricted subspace formed by a collection of such Majorana bound states, a quantum computer would perform calculations through braiding operations~\cite{Nayak2008}.

\comment{Semiconductor-superconductor heterostructures are the way to go.}
So far, several physical platforms have been proposed to realize Majorana bound states: topological insulator-superconductor~\cite{Fu2008}, semiconductor-superconductor (SM-SC) heterostructures~\cite{Oreg2010, Lutchyn2010, Sau2010}, or magnetic atom chains~\cite{Choy2011, Nadj-Perge2013, Braunecker2013, Klinovaja2013}.
The growing number of theoretical proposals and avenues investigated in experiments was surveyed in several recent reviews~\cite{Alicea2012, Leijnse2012, Beenakker2013, Sarma2015, Elliott2015, Sato2016, Aguado2017}.

\comment{Different ways to detect the topological transition in heterostructures}
The focus in our paper is on the, maybe, most promising condensed matter candidate for the realization of Majorana bound states: the SM-SC heterostructures~\cite{Stanescu2013, Lutchyn2018, Zhang2019}.
The first experimental signatures of Majorana bound states were obtained by measuring a zero-bias peak in the tunneling conductance~\cite{Mourik2012}.
Other proposed measurements are to detect Majoranas using the fractional Josephson effect~\cite{Kitaev2001, Kwon2003, Kwon2004, Fu2009, Jiang2011}, and in current correlations~\cite{Bolech2007, Nilsson2008, Law2009, Golub2011}.
The detection of Majorana states remains still open to debate as the signal sought from them may be due to low-energy Andreev bound states trapped in the hetereostructure due to smooth confining potentials~\cite{Kells2012, Prada2012, Cayao2015, Liu2017, Moore2018, Moore2018a, Awoga2019, Vuik2019}.

\begin{figure}[t]
\includegraphics[width=\columnwidth]{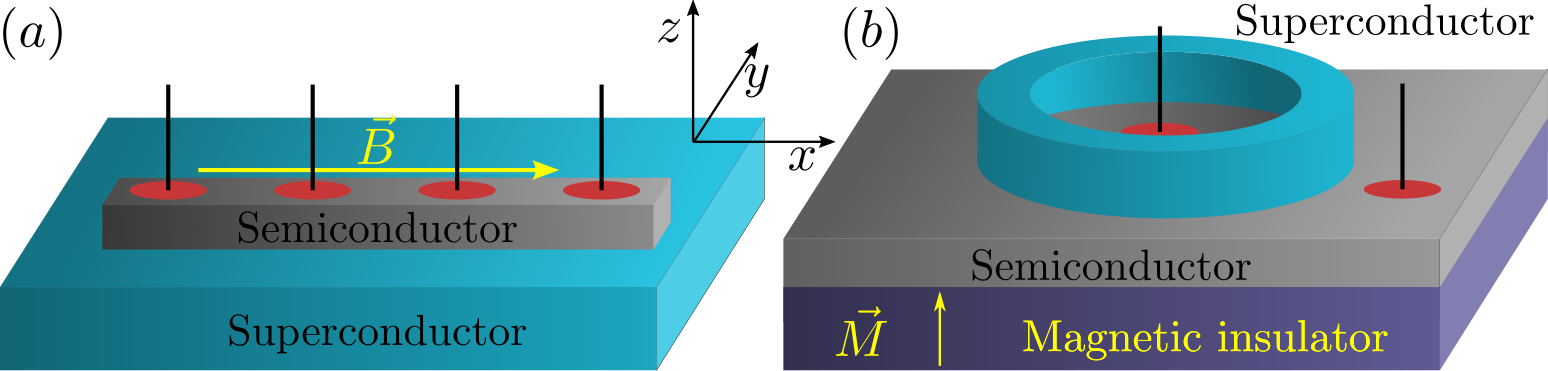}
\caption{The present paper studies the (a) 1D and the (b) 2D models proposed in Refs.~\onlinecite{Lutchyn2010, Oreg2010}, respectively, Ref.~\onlinecite{Sau2010}, as physical platforms supporting Majorana bound states. 
Additional gates with time-oscillating voltage are attached to the semiconductor (in red), allowing to modulate the spin-orbit coupling.
The response of the systems to the alternating electrical field discriminates near the topological transition the nontrivial phases.}
\label{fig:devices}
\end{figure}

\comment{Indirect detection}
Another fruitful alternative is to detect the topological phases in SM-SC heterostructures by indirect means, using, for example, bulk measurements.
Signatures of the topological transitions have been theoretically shown to arise in the electromagnetic response of the system to weak time-dependent magnetic fields~\cite{Ojanen2012, Ojanen2013}, in the entanglement spectrum of $p$-wave superconductors~\cite{Oliveira2014}, in dynamical probes of one-dimensional ultracold atomic gases with Majorana modes~\cite{Setiawan2015}, in critical currents~\cite{Cayao2017} and nonlocal conductance measurements~\cite{Rosdahl2018} of Josephson junctions, and so on.
Recently it was realized that at the topological phase transition the spin polarization of electronic bands is inverted, a feature that might be exploited as a reliable marker to discriminate the topological phases~\cite{Szumniak2017}.
Further studies have sought to make use of this observation to devise detection methods using the local measurement of spin in the electronic bands at the transition~\cite{Serina2018}, in the generation of supercurrents~\cite{Chen2019}, or using spin-selective measurements via quantum dots connected to the heterostructure~\cite{Chevallier2018, Juenger2019}.
In this paper, we propose an alternative detection method which relies on all-electrical probes of the system's bulk electronic structure, coupled with optical detection.

\comment{Our systems}
Our analysis carries on two SM-SC heterostructures, where the existence of Majorana phases has been proposed---a one-dimensional model~\cite{Lutchyn2010, Oreg2010} and a two-dimensional one~\cite{Sau2010} (see Fig.~\ref{fig:devices}).
Both setups aim to realize an effective spinless $p$-wave superconductor with topological properties~\cite{Kitaev2001} in the low-energy sector, near the Fermi energy.
The basic ingredients are a magnetic field, which removes the Kramers degeneracy of the electronic states, while a strong spin-orbit coupling breaks the spin conservation to allow tunneling of Cooper pairs from a neighboring $s$-wave superconductor into the semiconductor.
This induces a superconducting gap in the semiconductor, creating an effective topological superconductor.
The magnetic field modifies the spectrum, acting against the induced superconducting gap, allowing to close the spectral gap for a critical magnetic field, at zero wave vector in the proximitized semiconductor.
Above the critical field, the gap reopens, and the system enters a topological nontrivial phase where Majorana bound states are expected to form.

\begin{figure}[t]
\includegraphics[width=\columnwidth]{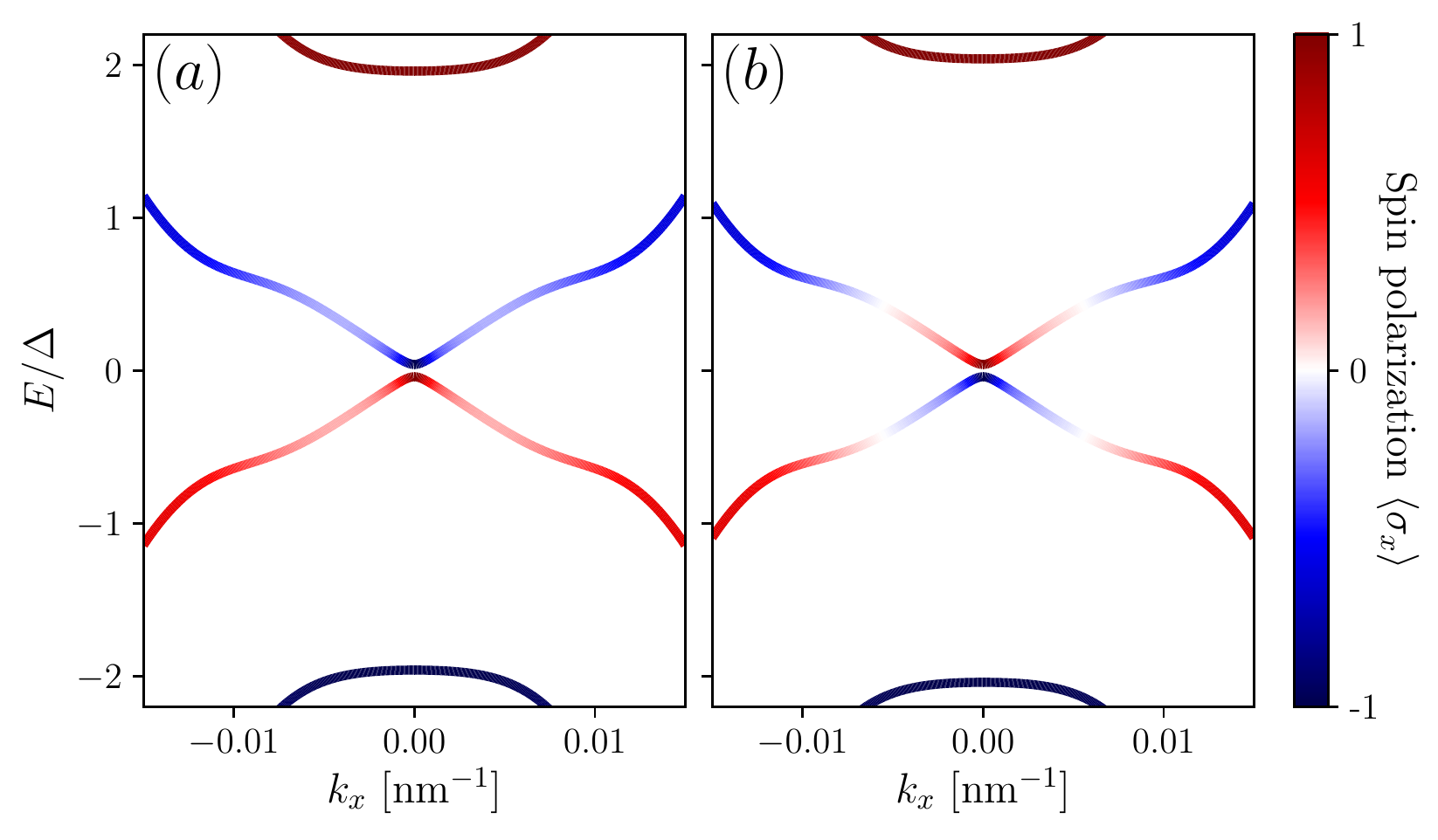}
\caption{Spin polarization $\avg{\sigma_x}$ of the electronic bands of the 1D model, close to the topological transition, in the (a) trivial phase and in the (b) nontrivial phase. The spectral gap has the same magnitude in both cases, $|E_g|=0.04\Delta$ and $\mu=0$.}
\label{fig:polariz}
\end{figure}

\comment{Spin inversion}
This basic physical picture readily allows one to understand the band spin inversion at the topological transition~\cite{Szumniak2017}.
At the zero wave vector, near the topological transition, the spin-orbit term is dominated by the Zeeman field which sets the band spin orientation either parallel or antiparallel to it.
Due to low-energy particle-hole symmetry, opposite-energy quasiparticle bands have opposite spin polarization.
Since at the topological transition the gap closes and the bands cross each other, while remaining spin polarized, the spin polarization of these bands is inverted between the trivial and nontrivial phase (see, e.g., Fig.~\ref{fig:polariz}).
This picture is limited to a region around $k=0$, since spin-orbit coupling acts at finite momenta to rotate the electronic spins.
The challenge of the present paper is to find signatures of the electronic spin inversion at the transition.

\comment{All-electrical detection}
We propose an all-electrical detection of spin polarization in the electronic bands in semiconductor-superconductor heterostructures which is capable of discriminating the phases near the topological transition.
Our proposal is to use techniques similar to the electronic spin resonance (ESR) spectroscopy, where  spin relaxation is measured using microwave-frequency magnetic fields~\cite{Slichter1990}.
However, in the SM-SC heterostructures, the proximity to a superconductor renders such methods not ideal.
The present all-electrical scheme relies on the indirect coupling of the electric field to the electronic spin due to the strong spin-orbit coupling present in semiconductors such as~\ce{InSb} or \ce{InAs}, which are regularly used in building the SM-SC heterostructures.
The electric fields have been shown to control the Land\'e $g$-factor in semiconductor devices~\cite{Nitta1997} and, moreover, time-varying electric field may be used to dynamically modulate the $g$-factors as a means to control quantum spins~\cite{Salis2001, Kato2003, Kato2004, Tang2006, Nowack2007, Jeff2014, Pawlowski2014, Pawlowski2016}.
Moreover, a time-dependent spin-orbit coupling has been predicted to generate spin currents~\cite{Malshukov2003, Tang2005, Liang2009, Ho2014}.

\comment{Our proposal}
In our proposal, the electric fields modulate the strength of the Rashba spin-orbit coupling in the material.
Then, under the electric field, resonant transitions are induced between the low-energy quasiparticle bands, leading to an increase in the spin polarization in either trivial or nontrivial phases.
Nevertheless, since the spin polarization is opposite in the two phases, longitudinal spin-relaxation processes near $k=0$ are either favored or unfavored by the external magnetic field.
We show that in the topological nontrivial phase the quasi-electrons have spins aligned with the magnetic field, and therefore they relax by emitting photons, while in the trivial phase, they relax by absorbing photons.
This allows the use of optical spectroscopic probes to detect the topological phases.
The associated response function $\chi(\omega)$, defined below in Eq.~\eqref{eq:chi} and in particular its imaginary part $\chi''(\omega)$, which is related to spin relaxation processes, encodes these features and distinguishes on which side of the topological transition is the system.
We call such measurement Rashba spectroscopy, sharing ideas from a larger group of experimental methods developed under the name of $g$-tensor modulated resonance spectroscopy~\cite{Kato2003}.

\comment{Organization of the paper}
The paper is organized as follows.
Section~\ref{sec:models} introduces the two Hamiltonian models for the semiconductor-superconductor heterostructures, in one and two dimensions.
Section~\ref{sec:rashba_spec} discusses the detection of band spin polarization using electrical modulation of the Rashba spin-orbit coupling. 
The section defines a response function modeling the experiment and is further determined within this paper.
The section closes with a discussion about the trivial effect of electric field coupling to the electronic density in the topological superconductor.
A more detailed treatment of this effect is relegated to Appendix~\ref{sec:dens_resp}. 
Sec.~\ref{sec:mu0} analyzes the system response at vanishing chemical potential, where a complete analytical solution is available.
The results are extended in the next Sec.~\ref{sec:pert_alpha} in a perturbation theory in small spin-orbit coupling near the topological transition.
The perturbation theory yields also the spin polarization of the electronic bands.
Section~\ref{sec:gen_sol} generalizes the above results for any system parameters in the low-frequency regime, while an arbitrary frequency formula for the response function is relegated to Appendix~\ref{sec:response}.
Section~\ref{sec:finite_disorder_systems} verifies the robustness of the Rashba susceptibility signal in finite tight-binding models with potential disorder.
Finally, Sec.~\ref{sec:conc} sums up the conclusions of our study.

\section{Models}
\label{sec:models}
In this paper we investigate two paradigmatic models, originally 
proposed in Refs.~\onlinecite{Lutchyn2010, Oreg2010, Sau2010}, as condensed matter platforms for the realization of Majorana bound states.
Because of its relative simplicity, the one-dimensional (1D) model has been the subject of intense experimental scrutiny~\cite{Mourik2012,Deng2012,Rokhinson2012,Das2012,Churchill2013,Albrecht2016,Guel2018,Zhang2018}.
To treat the models on equal footing, we assume in both cases that the semiconductor is deposited on a superconductor in $xy$ plane (see Fig.~\ref{fig:devices}).
The proximity to the superconductor induces superconducting correlations in the semiconductor, characterized by the order parameter $\Delta$.
The semiconductors are also characterized by a strong Rashba spin-orbit coupling due to broken inversion symmetry along the $z$ axis.
Finally, the time-reversal symmetry is broken by an (effective) magnetic field which gives rise to a Zeeman spin-splitting between the electronic bands in the semiconductor.
In the 1D model, the external magnetic field is applied along the semiconducting wire.
In the two-dimensional (2D) setup there is only an effective Zeeman field  perpendicular to the semiconducting plane induced by a magnetic insulator placed under the semiconductor.

The effective Hamiltonian for both semiconductors reads
\begin{equation}
\mathcal H = \frac{1}{2} \int d^d\bm k \Psi^\dag(\bm k) H(\bm k) \Psi(\bm k),
\end{equation}
with the Nambu field operator defined as $\Psi^\dag(\bm k)=(\psi_{\bm k\uparrow}^\dag,\psi_{\bm k\downarrow}^\dag, \psi_{-\bm k\downarrow},-\psi_{-\bm k\uparrow})$.
We use the convention that
\begin{equation}
\begin{aligned}
\bm k&=(k_x,0,0),   && k=k_x  && (\rm 1D), \notag \\
\bm k&=(k_x,k_y,0), && k=(k_x^2+k_y^2)^{1/2} && (\rm 2D)\notag.\\
\end{aligned}
\end{equation}

The Bogoliubov--de Gennes Hamiltonian for the 1D model reads~\cite{Lutchyn2010, Oreg2010}
\begin{equation}\label{1d_ham}
H(k_x) = \Big(\frac{\hbar^2k^2_x}{2m}-\mu
\Big)\tau_z
-\alpha\, k_x\tau_z\sigma_y+E_{\rm Z}\,\sigma_x+\Delta\tau_x,
\end{equation}
where, without the loss of generality, we choose an uniform induced order parameter $\Delta>0$.

In the 2D model, the order parameter $\Delta$ has a vortex structure, and goes to 0 in the middle of the annular structure shown in Fig.~\ref{fig:devices}(b).
Our focus is on the bulk excitation spectrum which is determined far away from the vortex, where the order parameter is assumed to have an uniform amplitude $\Delta>0$.
The Hamiltonian for the system under the above approximation reads~\cite{Sato2009,Sau2010}
\begin{eqnarray}\label{2d_ham}
H(\bm k) &=& \Big(\frac{\hbar^2 k^2}{2m}-\mu
\Big)\tau_z
+H_{\rm so}+E_{\rm Z}\,\sigma_z+\,\Delta\tau_x,\notag\\
H_{\rm so}(\bm k)&=&\alpha\,\tau_z(\sigma_xk_y-\sigma_yk_x).
\end{eqnarray}
In both models, $E_{\rm Z}$ denotes the Zeeman energy, $\alpha$, the Rashba spin-orbit coupling strength, and $\mu$ the chemical potential. 
The Pauli matrices $\sigma_i$ act in spin space, while $\tau_i$, with $i=x,y,z$, in particle-hole space. 
We use the convention that $\tau_i\sigma_j \equiv \tau_i\otimes\sigma_j$, and the absence of a Pauli matrix in the Hamiltonian implies the presence of the identity matrix in the respective space.

Despite the somewhat different physical realization, the models share many attributes, allowing throughout a parallel treatment and leading to similar conclusions.
Formally, the 2D model reduces to the 1D model under a rotation in spin space and confinement of electron motion along the $x$ axis.
Since the Rashba spin-orbit vector is orthogonal to the effective magnetic field, the energy spectrum is determined analytically.
In both models there are two positive-energy quasiparticle bands
\begin{eqnarray}\label{eig_energies}
E_\pm(k) &=& [\xi^2+\alpha^2k^2+E_{\rm Z}^2+\Delta^2\notag\\
&&\pm 2(\alpha^2k^2\xi^2+E_{\rm Z}^2\xi^2+E_{\rm Z}^2\Delta^2)^{\frac12}]^{\frac12},
\end{eqnarray}
with two negative-energy bands $-E_{\pm}(k)$, with $\xi$ denoting the kinetic energy $\xi=\hbar^2k^2/2m-\mu$.
The band structures undergo a topological transition when the spectral gap at $k=0$, 
\begin{equation}\label{delta_top}
E_g = E_{\rm Z} - \sqrt{\Delta^2+\mu^2},
\end{equation}
closes and reopens under a variation of system parameters.
The topological nontrivial phases are realized for $E_g>0$ with zero energy Majorana bound states localized either at the 1D wire edges or, in 2D model, in the superconductor vortex.

Near the topological phase transition at $k=0$, the spin-orbit coupling term is dominated by the Zeeman field which polarizes the quasiparticle bands parallel or antiparallel to it. 
Due to particle-hole symmetry, bands with opposite energies display opposite spin polarization.  
A more detailed discussion of the spin polarization near the transition is presented in Sec.~\ref{sec:pert_alpha}.

To get a sense of the units involved, we take throughout an \ce{InSb} semiconductor with the material parameters~\cite{Mourik2012}: $g$-factor $\sim 50$, effective mass $\sim 0.015\,m_e$, $\alpha=\SI{20}{nm\cdot meV}$, and induced superconducting gap $\Delta=\SI{0.25}{meV}$.
We investigate systems that exhibit spectral gaps on the order of $E_g \sim \SI{0.05}{meV}$, which puts the frequency in the range $\omega\sim \SI{75}{GHz}$.
Therefore the systems could be probed in the microwave regime.
For the sake of simplicity we take throughout a similar set of parameters in the 2D model and present all energies in units of $\Delta$.

\section{Rashba spectroscopy and the response function}
\label{sec:rashba_spec}
To probe the system, a time-dependent electric field is generated perpendicular to the superconductor $\delta E(t)\hat{\bm z}$, for example, by laser pulses and microwaves exciting a voltage gate connected to the proximitized semiconductor~\cite{Kato2003, Nowack2007}.
Alternatively, one can imagine modulating a perpendicular electric field applied directly to the system. 
The electric field generates an effective in-plane magnetic field which couples with the spins in the semiconductor.
This yields a time-dependent modulation of the Rashba spin-orbit coupling of the form,
\begin{equation}
\delta H_{\rm so}(t) = \tau_z(\bm\sigma\times\bm k)\cdot\hat{\bm z}\delta \alpha(t) = \frac{H_{\rm so}}{\alpha}\delta \alpha(t).
\end{equation}
In general, since the spin-orbit coupling strength depends linearly on the external electric field in Rasbha nanowires~\cite{Wojcik2018}, its time modulation remains linear in the electric field, $\delta\alpha(t) \simeq \kappa \delta E(t)$.
This allows us to investigate the system using perturbative approaches.

 
Near the topological transition, i.e., $E_g=0$, the effect of spin-orbit coupling is small and the Zeeman field polarizes the quasiparticle bands along its direction.
The time-dependent perturbation creates quasiparticle excitations, which change their spin polarization.
In linear response theory, the change in the polarization
\begin{equation}\label{rashba_spec}
\delta\langle\sigma_{j}(t)\rangle
=\int_{-\infty}^t \chi_{jR} (t-t')\delta\alpha(t')dt',
\end{equation}
is measured by the susceptibility $\chi_{jR}(t)$, with $j\in \{x, y, z\}$.
More exactly, the response function $\chi_{jR}(t)$ measures the indirect coupling of the external electric field to the electronic spin $\sigma_j$ due to the strong Rashba coupling  present in the semiconductor.
The expectation values are computed in a basis of the Hamiltonian eigenstates $H(\bm k)|n\bm k\rangle = E_{n}(k)|n\bm k\rangle$, with $n$ a band index.
The spin polarization of a $\bm k$ state is therefore denoted $\avg{\sigma_j}\equiv\avg{n\bm k|\sigma_j|n\bm k}$.

The energy provided by the electric field causes resonant optical (momentum-conserved) transitions for electrons between the quasiparticle bands, and it should be in the microwave range, according to our estimates.
These transitions are detected in what we call Rashba spectroscopy, by some of the components of $\chi_{jR}(\omega)$, in analogy to the ESR spectroscopy.
The latter involves measuring spin-spin correlations functions, due to direct coupling of external ac-magnetic fields to the electronic spin. 
In contrast, Rashba spectroscopy measures the response caused by the coupling between the modulated Rashba spin-orbit term to the electronic spin. 
Therefore the dynamical long-wavelength response function reads:
\begin{equation}
\chi_{jR} (t-t') = -i\theta(t-t')
\langle
[\sigma_{j}(t),\tau_z(\bm \sigma(t')\times \bm k)\cdot \hat{\bm z})]
\rangle,\label{eq:chi}
\end{equation}
with $\bm k$ defined accordingly either to the 1D or the 2D model, and $[\cdot,\cdot]$ is a commutator.
Similar ideas, in the context of quantum dot spin control in semiconductor quantum wells, have been experimentally put forward under the name of $g$-tensor modulation resonance spectroscopy~\cite{Kato2003}.

The response function is invariant under time translations, and therefore a Fourier transform yields readily its frequency-dependent expression
\begin{eqnarray}\label{susc}
\chi_{jR}(\omega)&=&\sum_{mn}\int\frac{d^d\bm k}{(2\pi)^d}
[f_{m}(k)-f_{n}(k)]\\
&&\times\frac{\langle m\bm k|\sigma_{j}| n\bm k\rangle 
\langle n\bm k|H_{\rm so}/\alpha |m\bm k \rangle}
{\hbar\omega+E_{m}(k)-E_{n}(k)+i\delta}\notag
\end{eqnarray}
with $\delta/\Delta\to 0^+$.
The summation is over the four quasiparticle bands, and the momentum integration carries over the available momentum states.
The Fermi-Dirac function is $f_{n}(k)=[e^{\beta E_{n}(k)}+1]^{-1}$, with $\beta=1/k_BT$, the inverse temperature.

Alternatively, the dynamical correlator $\chi_{jR}(\omega)$ is calculated within the Matsubara Green's function formalism
\begin{equation}\label{chi_mat}
\chi_{jR}(i\omega) = \frac{1}{\beta}\sum_{\nu}\int \frac{d^d\bm k}{(2\pi)^{d}} 
\text{Tr}[\sigma_j G(\bm k, i\nu) \frac{H_{\rm so}}{\alpha} G(\bm k, i\omega+i\nu)],
\end{equation}
with $G(\bm k, i\nu) = 1/(i\nu-H)$, the superconducting Green's function at fermionic Matsubara frequencies $\nu$.

While the response function is in general complex, $\chi=\chi'+i\chi''$, we focus on its imaginary part $\chi''(\omega)$, which carries information about the spin relaxation processes, and is sensitive to the spin polarization of the quasiparticle bands.
In particular, we show that only the components of $\chi_{jR}''(\omega)$ along the Zeeman field discriminate between trivial and nontrivial superconducting phases at the topological transition as the susceptibility changes sign across the transition at $E_g=0$. 
According to the effective magnetic field orientation chosen in Eqs.~\eqref{1d_ham} and~\eqref{2d_ham}, the relevant susceptibilities are denoted
\begin{equation}
\chi_{1}(\omega)\equiv\chi_{xR}(\omega)\quad\text{(1D)},\quad
\chi_{2}(\omega)\equiv\chi_{zR}(\omega)\quad\text{(2D)}.
\label{eq:susceptibility}
\end{equation}
The other components are vanishingly small near the phase transition, for frequencies on the scale of the spectral gap $\hbar\omega\sim 2|E_g|$, since, in this limit, the bands are almost completely polarized by the Zeeman field.

Finally, there is an additional contribution to the change in spin polarization which is due to the coupling between the time-varying electric field and the electronic density in the semiconductor.
The effect of the electric field is described at the linear response level by the susceptibility
\begin{equation}
\chi_{j\mu}(t-t')=-i\theta(t-t')\avg{[\sigma_j(t),-\tau_z(t')]}.
\end{equation}
This component to the response is trivial since its imaginary part does not change sign across the topological transition.
We treat in detail this susceptibility in Appendix~\ref{sec:dens_resp}.
It is noteworthy to briefly go over some of its properties. 

The response $\chi_{j\mu}$ is nonzero only for spins in the direction of the magnetic field, as in the case of Rashba susceptibility.
The excitation threshold is not modified and $\chi''_{j\mu}$ becomes nonzero only for frequencies $\hbar\omega>2|E_g|$. This is 
expected since in linear response the spectrum of the Hamiltonian is not modified by the fluctuating chemical potential.

We can profit from the fact that the two components to the susceptibility have qualitatively different behavior at the topological transition and extract the
Rashba response even without accounting for their magnitude.
While Rashba susceptibility $\chi''_{jR}$ is odd in $E_g$, the electronic density component, $\chi''_{j\mu}$ is even in $E_g$.
Therefore, it is possible to discriminate the two signals by measuring the total susceptibility for different gaps, symmetrically around $E_g=0$, and eliminate the trivial response by doing symmetric and antisymmetric combinations of the recorded signal.
In the following, we focus on the Rashba susceptibility since it is the one encoding the nontrivial response of the system to the electric field.

\section{Analytical solution for \texorpdfstring{$\mu=0$}{mu=0}}
\label{sec:mu0}
To analyze the response function $\chi_d(\omega)$,  it is useful to investigate the limit $\mu=0$ where closed-form solutions are possible.
Later, we demonstrate that the main features captured in this limit carry over to more general choices of parameters.

Let us focus on the low-energy physics near the Fermi energy at $E=0$.
The lower band $E_-(k)$, given in Eq.~\eqref{eig_energies}, displays minima both at $k=0$ and $k\simeq 2m\alpha/\hbar^2$ (for strong spin-orbit strength).
The induced superconducting correlations open a superconducting gap $\sim \Delta$ at finite momenta $\bm k$.
In contrast, the spectrum at $k=0$ is defined by the gap Eq.~\eqref{delta_top}.
The closing and reopening of $E_g$ marks a transition from the topological trivial phase ($E_g<0$) to the nontrivial phase supporting Majorana bound states $(E_g>0)$.
Our analysis is concerned in the parameter regime around the phase transition point, where $|E_g|\ll\Delta$.
Under this approximation only momenta near $\bm k=0$ are relevant and quadratic terms in momentum are neglected.
Moreover, we work at vanishing chemical potential, $\mu=0$, and therefore $\xi\simeq 0$ and 
\begin{equation}
E_g=E_{\rm Z}-\Delta.
\end{equation}

At $\bm k=0$, the lowest-energy band and its particle-hole partner are eigenstates of $\sigma_x$ in the 1D model, and $\sigma_z$ in the 2D one. 
Due to particle-hole symmetry the two bands have opposite polarization $\avg{\sigma_j}$ (see Fig.~\ref{fig:polariz}).
At the transition point the two eigenstates cross and there is a change in the polarization of the crossing bands. 
This change in polarization is detected by the imaginary part of the response function $\chi_d(\omega)$, defined in Eq.~\eqref{eq:susceptibility}.

\begin{figure}[t]
\includegraphics[width=\columnwidth]{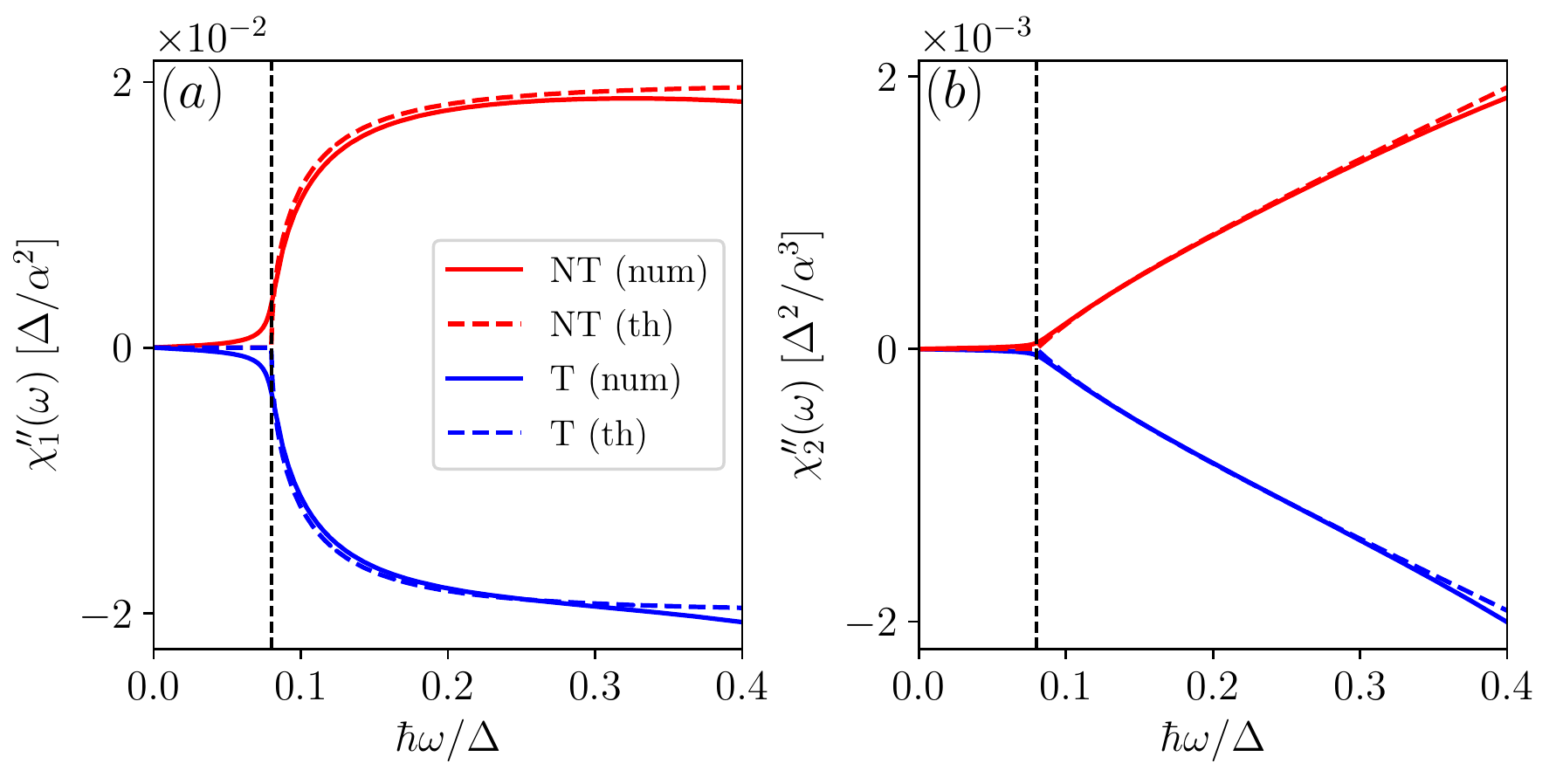}
\caption{Imaginary part of the susceptibility $\chi_{d}''(\omega)$ as a function of frequency for (a) the 1D model and (b) the 2D model.
The spectral gap has the same magnitude in all cases $|E_g|=0.04\Delta$, but, according to the sign of $E_g$, the systems may be either in a topologically trivial phase (T, blue), or a nontrivial one (NT, red). 
The response shows different behavior depending on the phase.
The solid lines are obtained by numerical integration in Eq.~\eqref{chi_mat}, while the dashed lines plot the analytical results in Eqs.~\eqref{imag_chi_1d_th} and~\eqref{imag_chi_2d_th}.
The panels share the legend.
}
\label{fig:img_chi_mu0}
\end{figure}

In the zero-temperature limit, the sum over the Matsubara frequencies $\nu$ in Eq.~\eqref{chi_mat} may be replaced by an integral. 
The susceptibility $\chi_d(\omega)$ for both 1D and 2D models follows after performing the trace over particle-hole and spin degrees of freedom,
\begin{eqnarray}
\chi_d(i\omega) &=& \int \frac{d^d\bm k d\nu}{(2\pi)^{d+1}} \frac{4\alpha k^2
E_g}{[E_-(k)^2+\nu^2][E_-(k)^2+(\omega+\nu)^2]}\notag\\
&&+\,(\Delta\to-\Delta),\label{eq:integral}
\end{eqnarray}
with
$E_-(k) = (\alpha^2 k^2+E_g^2)^{1/2}.$
The second term in $\chi_d(\omega)$ contributes to the imaginary part of the susceptibility only at higher frequency, equal or larger than the separation between the lowest and highest bands $\sim\!2(E_{\rm Z}+\Delta)$.
Therefore it can be neglected when probing the system at smaller frequencies, $\hbar\omega\sim 2|E_g|$.
The first term in Eq.~\eqref{eq:integral} gives the low-frequency contribution which is, as expected, proportional to $E_g$, and, furthermore, is changing sign at the topological transition.
We note that, in contrast, the static susceptibility $\chi_d'(\omega=0)\propto\int d\omega' \chi''_d(\omega')/\omega'$ is an unreliable marker of the topological transition, since it includes the information from the high-frequency transitions.


In the 1D model, the low-frequency dynamical susceptibility for transitions between the low-energy bands follows after performing the integral over the Matsubara frequency $\nu$ and the analytical continuation $i\omega\to \omega+i\delta/\hbar$:
\begin{equation}\label{chi_poles_1}
\chi_{1}(\omega)=\int \frac{dk k^2}{\pi E_-(k)}
\bigg[
\frac{2\alpha E_g}{4E_-(k)^2-(\hbar\omega+i\delta)^2}
\bigg],
\end{equation}
whose imaginary part is 
\begin{equation}\label{imag_chi_1d_th}
\chi''_{1}(\omega)=\frac{E_g}{\alpha^2\hbar\omega}\sqrt{\frac{\hbar^2\omega^2}{4}-E_g^2}
\Theta\bigg(\frac{\hbar^2\omega^2}{4}-E_g^2\bigg),
\end{equation}
where $\Theta$ is the Heaviside step function. 

\begin{figure}[t]
\includegraphics[width=\columnwidth]{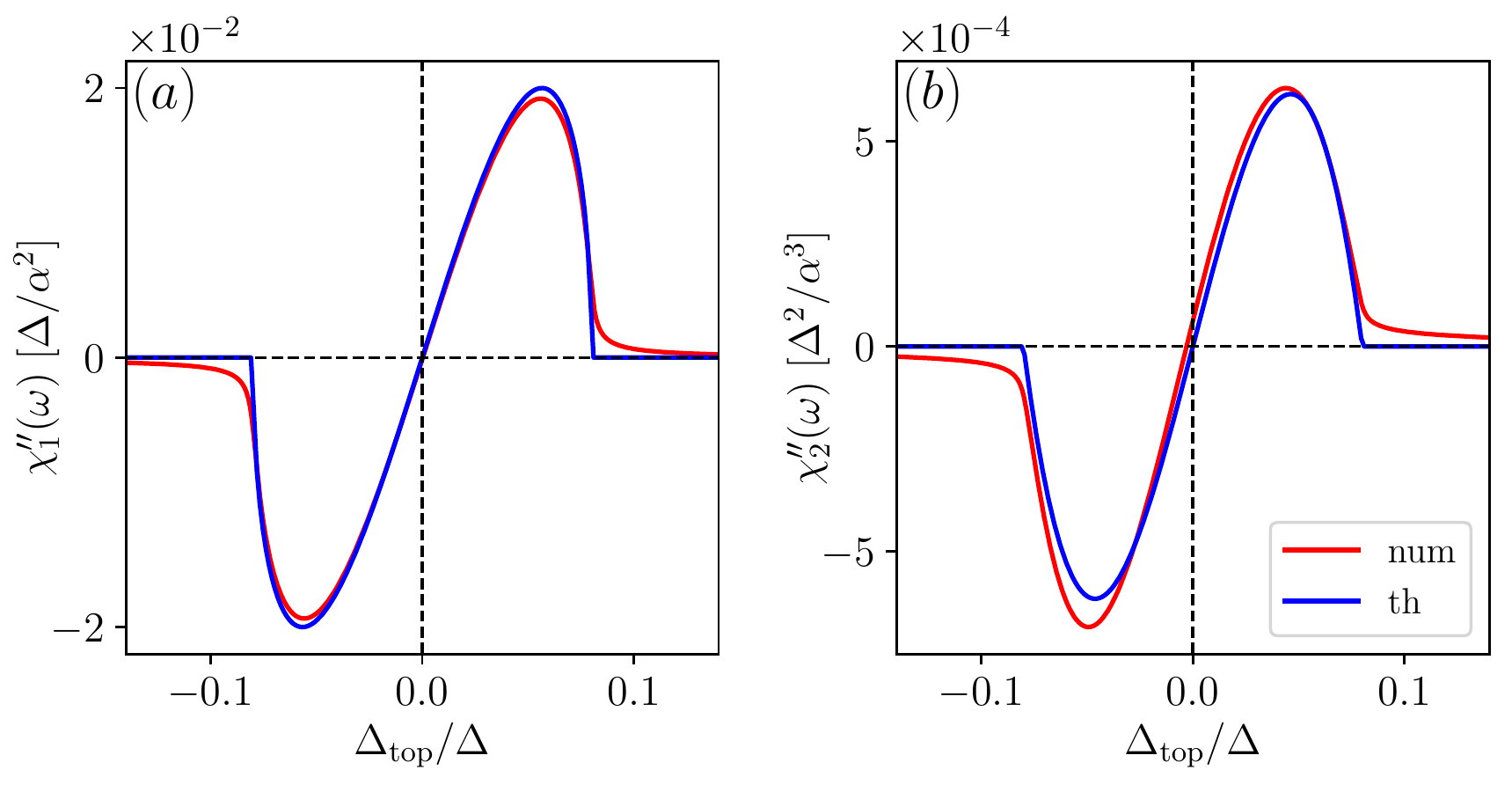}
\caption{Imaginary part of the susceptibility $\chi_{d}''(\omega)$ as a function of the spectral gap $E_g=E_{\rm Z}-\sqrt{\Delta^2+\mu^2}$ for (a) the 1D model and (b) the 2D model.
The frequency is fixed at $\hbar\omega=0.16\Delta$.
The susceptibility is computed either from the analytical expressions~\eqref{imag_chi_1d_th} and~\eqref{imag_chi_2d_th} (red line), or numerically from Eq.~\eqref{chi_mat} (blue line).
The panels share the legend.
}
\label{fig:trans_ez}
\end{figure}

In the 2D model, an additional trivial angular integration in the 2D plane is required, which yields
\begin{equation}\label{chi_poles_2}
\chi_{2}(\omega) =
\int_0^{\infty} 
\frac{dk k^3}{\pi E_-(k)}
\bigg[
\frac{2\alpha E_g}{4E_-(k)^2-(\hbar\omega+i\delta)^2}
\bigg].
\end{equation}
Therefore the imaginary susceptibility reads
\begin{equation}\label{imag_chi_2d_th}
\chi''_{2}(\omega)=\frac{E_g}{2\alpha^3 \hbar\omega}
\bigg(\frac{\hbar^2\omega^2}{4}-E_g^2\bigg)
\Theta\bigg(\frac{\hbar^2\omega^2}{4}-E_g^2\bigg).
\end{equation}
As expected,  $\chi_d''(\omega)$ is odd in frequency and, due to vanishing density of states in the spectral gap, is zero below the gap.
At a threshold $2|E_g|$, which is the energy gap between the lowest bands $\pm E_-(k)$, the 1D response develops a square-root dependence on the frequency, while the 2D susceptibility displays a linear dependence.
A comparison between the analytical predictions and numerical integration of $\chi_d$ either using Eq.~\eqref{susc} or~\eqref{chi_mat} is presented in Fig.~\ref{fig:img_chi_mu0}.

The response functions change sign at the topological transition, an observation that can be validated experimentally.
Moreover, in experiments it is also possible to keep the frequency fixed, but to vary the Zeeman field to bring the system across the topological transition.
Near the transition at $E_g=0$ the response is linear in $E_g$ as indicated by Eqs.~\eqref{imag_chi_1d_th} and~\eqref{imag_chi_2d_th}. 
At larger Zeeman field, above the fixed electric-field frequency $\omega$, the transitions between the bands are energetically unfavored, leading to a decay of the signal.
The dependence of $\chi''_d(\omega)$ on $E_g$, when increasing $E_{\rm Z}$, is displayed in Fig.~\ref{fig:trans_ez}, showing the expected sign change at the transition.
We also note that with increasing frequency, additional transitions to higher-energy bands are also possible, but the low-frequency response close to the topological transition is insensitive to them.

\section{Perturbation theory in the spin-orbit coupling}
\label{sec:pert_alpha}
The results of the previous section are extended here to finite chemical potential $\mu$ using a perturbation theory in the spin-orbit coupling strength near the topological transition. 
This allows an intuitive understanding of the processes modeled in the Rashba spectroscopy response function.

The perturbation theory is justified close to the topological transition at $\bm k=0$ where the spin-orbit coupling term, which is linear in momentum, is dominated by the other terms in the Hamiltonian.
The kinetic term $\sim k^2/2m$ remains neglected, since it is quadratic in momentum.

To be more specific, in this section, we focus on the 1D model, described by a simplified Hamiltonian
\begin{eqnarray}
H &=& H_0 + H_{\rm so}, \\
H_0&=&-\mu\tau_z+E_{\rm Z}\sigma_x+\Delta\tau_x,\notag
\end{eqnarray}
with the Rashba spin-orbit term $H_{\rm so}=-\alpha k_x\tau_z\sigma_y$ as a perturbation on $H_0$.

Our goal is to determine $\chi_1(\omega)\equiv\chi_{xR} (\omega)$, proving that it changes sign at the topological transition $E_g=0$, with
\begin{equation}
E_g = E_{\rm Z} - \sqrt{\Delta^2+\mu^2}.
\end{equation}
The response function~\eqref{susc} in the zero-temperature limit follows readily using the eigenstates of the Hamiltonian, determined within the perturbation theory.

Let us perform a $\pi/2$-rotation around $x$ axis in particle-hole space
\begin{equation}
\tau_z\mapsto \tau_y,\quad \tau_y\mapsto -\tau_z,
\end{equation}	
only for notational simplicity. The Hamiltonian changes accordingly, $H\mapsto \tilde H$, with the tilde denoting the effect of the unitary transformation.
 
The $\tilde H_0$ eigenstates are momentum independent and may be indexed as $|\tau\sigma\rangle$ with $\tau=\pm$ and $\sigma=\pm$.
Since $|\tau\sigma\rangle$ are eigenstates of $\sigma_x$, it follows immediately that correlations $\chi_{zR}=\chi_{yR}=0$ and only the response along the magnetic field may be relevant.
The four energy bands of either $H_0$ or $\tilde H_0$ are
\begin{equation}
\varepsilon_{\tau\sigma} = \sigma E_{\rm Z}+\tau\sqrt{\Delta^2+\mu^2},
\end{equation}
with normalized eigenstates $|\tau\sigma\rangle$:
\begin{equation}
|\!+\!\pm\rangle = \frac{1}{2}\pmat{1\\e^{i\theta}}\otimes\pmat{1\\\pm 1},\,
|\!-\!\pm\rangle = \frac{1}{2}\pmat{e^{-i\theta}\\-1}\otimes\pmat{1\\ \pm 1},
\end{equation}
and
\begin{equation}\label{exp_theta}
e^{i\theta} = \frac{\Delta-i\mu}{\sqrt{\Delta^2+\mu^2}}.
\end{equation}
At the topological transition $E_g=0$, the bands $|\!+\!-\rangle$ and $|\!-\!+\rangle$ cross each other.
Note that in the trivial phase $E_g<0$, the ``conduction'' bands are $|\!+\!\pm\rangle$, while $|\!-\!\pm\rangle$ are ``valence'' bands.

Let us analyze the matrix elements in the susceptibility from Eq.~\eqref{susc} using the first-order perturbed eigenstates, linear in $\alpha$,
\begin{equation}
|\tau\sigma^{(1)}\rangle=|\tau\sigma\rangle-\sum_{\tau'\sigma'\neq\tau\sigma}
\frac{\avg{\tau'\sigma'|\tilde H_{\rm so}|\tau\sigma}}{\varepsilon_{\tau'\sigma'}-\varepsilon_{\tau\sigma}}|\tau'\sigma'\rangle.
\label{eq:perturbation}
\end{equation}
To first order, the only finite matrix elements of $\tilde H_{\rm so}$ are those between the valence and the conduction bands.
The modulated Rashba term $\tilde H_{so}/\alpha$, which couples to the time-varying electric field, excites quasiparticles from the lower to the upper bands.
Its matrix elements, are to lowest order independent on $\alpha$,
\begin{eqnarray}\label{mat_2}
\avg{+\!+^{(1)}\!|-k\tau_y\sigma_y|\!-\!-^{(1)}}&\simeq& k e^{-i\theta}\cos\theta,\notag\\
\avg{+\!-^{(1)}\!|-k\tau_y\sigma_y|\!-\!+^{(1)}}&\simeq& -k e^{-i\theta}\cos\theta.
\end{eqnarray}
This leads in either topological phases to an increase in the spin polarization for the upper band.

Relaxation processes are determined by the matrix elements of spins along the Zeeman field. 
To linear order in the spin-orbit coupling, they are given by:
\begin{eqnarray}\label{mat_1}
\avg{-\!\pm^{(1)}\!|\sigma_x|\!+\!\pm^{(1)}}&\simeq&0,\notag\\
\avg{-\!-^{(1)}\!|\sigma_x|\!+\!+^{(1)}}&\simeq& - \frac{\alpha ke^{i\theta}\cos\theta}{E_{\rm Z}+\sqrt{\Delta^2+\mu^2}},\\
\avg{-\!+^{(1)}\!|\sigma_x|\!+\!-^{(1)}}&\simeq&  
\frac{\alpha ke^{i\theta}\cos\theta}{E_g}.\notag
\end{eqnarray}
The second matrix element in Eq.~\eqref{mat_1} describes transitions between the highest and 
lowest energy bands ($\varepsilon_{++}$ and $\varepsilon_{--}$). 
The corresponding transition frequency is on order of $2(E_{\rm Z}+\Delta)$, which is much larger than the gap $E_g$, and it is therefore irrelevant for our analysis. 
Here we focus on the last matrix element in Eq.~\eqref{mat_1}, which is relevant for transitions between the two quasiparticle bands closest to the Fermi energy since the associated transition frequency is on the order of $\hbar\omega \sim 2|E_g|$.
Crucially, the matrix element behaves as $1/E_g$, so it changes sign at the topological transition.
This central result shows that the relaxation processes are dependent on whether the quasiparticles in the lowest conduction band are aligned to the magnetic field, as in the topological nontrivial phase (for $\varepsilon_{-+}$), or antialigned, as in the trivial phase (for $\varepsilon_{+-}$).
Note also that the spin-spin correlation functions, which model the conventional ESR experiments, would have in the present case a dependence on the absolute value of the spectral gap $\sim 1/E_g^2$, and therefore cannot discriminate the topological phases.

The intraband terms in the susceptibility are neglected since they are all real and do not contribute to the imaginary susceptibility $\chi_{1}''(\omega)$.
The transitions between highest and lowest energy bands are also neglected since they occur for higher frequencies than the ones comparable to the spectral gap $E_g$.
Therefore $\chi_1''(\omega)$ at low frequencies is determined only by the energy difference between the two quasiparticle bands closest to the Fermi energy.
To lowest order in $\alpha$, in a second-order perturbation theory, the energy difference reads
\begin{equation}\label{energy_diff}
\varepsilon_{-+}^{(2)} - \varepsilon_{+-}^{(2)}\simeq 2E_g + \frac{(\alpha k\cos\theta)^2}{E_g}
+\frac{(\alpha k \sin\theta)^2}{E_{\rm Z}}.
\end{equation}
The last term in Eq.~\eqref{energy_diff} may also be neglected, since it barely shifts the transition frequency due to the large value of the Zeeman energy $E_{\rm Z}\gg |E_g|$ (and it vanishes at $\mu=0$ or $\sin\theta=0$).
Then, the energy difference reads
\begin{equation}\label{square_approx}
\varepsilon_{-+}^{(2)} - \varepsilon_{+-}^{(2)}\simeq 2\mathrm{sgn}(E_g)\sqrt{|E_g|^2+(\alpha k\cos\theta)^2}.
\end{equation}
Using Eqs.~(\ref{mat_2} and \ref{energy_diff}) in Eq.~\eqref{susc} yields the susceptibility 
\begin{eqnarray}\label{1d_imag_chi}
\chi''_{1}(\omega)&=&
\int \frac{dk \alpha (k\cos\theta)^2}{2E_g}
\delta\big(\hbar\omega-2|E_g|-\frac{(\alpha k \cos\theta)^2}{|E_g|}\big)\notag\\
&&
-\,(\omega\to-\omega).
\end{eqnarray}

Again, the overall dependence on the sign of $E_g$ indicates that the susceptibility is a reliable marker for the topological transition.
This result translates the fact that in the nontrivial phase excited quasi-electrons relax by emitting photons at frequencies comparable to the spectral gap since they are aligned with the effective magnetic field, while in the trivial region, they absorb photons, since they are antialigned with it.
Integrating over the momentum in Eq.~\eqref{1d_imag_chi} and using the definition for $\cos\theta$ from Eq.~\eqref{exp_theta} we obtain
%
\begin{eqnarray}
\chi''_{1}(\omega)&=&\mathrm{sgn}(\omega E_g)
\frac{\sqrt{|E_g|(\Delta^2+\mu^2)}}{2\alpha^2|\Delta|}
\sqrt{\hbar|\omega|-2|E_g|}\notag\\
&&\times\,\Theta(\hbar^2\omega^2-4E_g^2).
\end{eqnarray}
The approximation of Eq.~\eqref{square_approx} yields an alternative result for the susceptibility
\begin{eqnarray}\label{pert_2}
\chi''_{1}(\omega)&\simeq&\frac{\hbar\omega\sqrt{\Delta^2+\mu^2}}{4\alpha^2|\Delta| E_g}
\sqrt{\frac{\hbar^2\omega^2}{4}-E_g^2}\notag\\
&&\times\,\Theta(
\hbar^2\omega^2-4E_g^2),
\end{eqnarray}
which reduces to the previous one at small $\alpha$ and $\hbar\omega \sim 2|E_g|$.
The susceptibility is odd in frequency, and changes sign with the spectral gap $E_g$. 
A quick check shows that Eq.~\eqref{pert_2} recovers the $\mu=0$ case from Eq.~\eqref{imag_chi_1d_th}, near the transition, with a frequency $\hbar\omega\sim 2|E_g|$.
Both analytical and numerical calculations show again that in the 1D model the susceptibility has a square-root dependence on frequency $\chi_1''(\omega)\sim\sqrt{|\omega|}$ near the topological transition [see Fig.~\ref{fig:img_chi_mu0}(a)].

Although the complete perturbation theory of the 2D case is not performed here, a simple scaling analysis shows that $\chi''_2(\omega)$ has, in general, a linear dependence on frequency, similar to the $\mu=0$ case displayed in Fig.~\ref{fig:img_chi_mu0}(b).
This behavior is understood by noticing that the matrix elements of the spin $\sigma_z$ and spin-orbit term are linear in $k$. 
Moreover, the susceptibility has poles at momenta $k_0\sim\sqrt{|\omega|}$, and 
therefore from Eq.~\eqref{susc} it follows that $\chi''_d\sim k_0^{d-1} k_0^2/k_0= k_0^d$, where the dimensional effects ($d=1$ or $2$) enter only from the integral measure.
Then indeed, in the 2D model, $\chi''_2\sim k_0^2\sim|\omega|$, as in the $\mu=0$ case.

Finally, the perturbation theory also yields the spin polarization in the two low-energy bands near the topological transition:
\begin{equation}\label{spin_polarization}
\avg{-\!+^{(1)}\!|\sigma_x|\!-\!+^{(1)}} \simeq 
1-\frac{(\alpha k\cos\theta)^2}{4E_g^2} -\frac{(\alpha k\sin\theta)^2}{4E_{\rm Z}^2},
\end{equation}
which generalizes at finite $\mu$ the results in Ref.~\onlinecite{Szumniak2017}.
Due to particle-hole symmetry, the eigenstate with opposite energy and momentum have also opposite polarization, and, since the polarization is even in $k$, 
\begin{equation}
\avg{+\!-^{(1)}\!|\sigma_x|\!+\!-^{(1)}}=-\avg{-\!+^{(1)}\!|\sigma_x|\!-\!+^{(1)}}.
\end{equation}
The energy of the state $|-\!+^{(1)}\rangle$ is below the Fermi energy in trivial region $E_g<0$ and above in the nontrivial region $E_g>0$.
Therefore there is an inversion in band polarization at the transition, as seen in Fig.~\ref{fig:polariz}.

The energy scale where the spin polarization first vanishes in the band sets a natural scale for the frequencies that one may use to probe the system. 
At larger momenta, the spin-orbit starts to dominate and reverts the polarization, such that at higher 
frequencies, the susceptibility may show no sign change.
In the approximation that $E_{\rm Z}\gg E_g$, we use Eqs.~\eqref{square_approx} and~\eqref{spin_polarization} to estimate that a reasonable frequency window to probe the system is $\hbar|\omega|\lesssim 6|E_g|$. 

A few remarks are in order.
As the spectral gap in the system increases, non-linear effects distort the band structure and the band minimum is no longer guaranteed at $k=0$.
The bending of the electronic bands lowers the energy of higher momentum states (of opposite spin polarization compared to the same-band $k\sim 0$ states), thus diminishing or reversing again the spin polarization of the band in a frequency window characteristic for $2E_-(k=0)$.
Therefore the detection method proposed here is expected generally to work whenever the minimum of the band is at the $\Gamma$ point, with frequencies tuned near the resonance condition, or, in particular, if the system is close to transition ($E_{\rm Z},\Delta \gg|E_g|$), with frequencies $\hbar|\omega|\lesssim 6|E_g|$. 

\begin{figure}[t]
\includegraphics[width=\columnwidth]{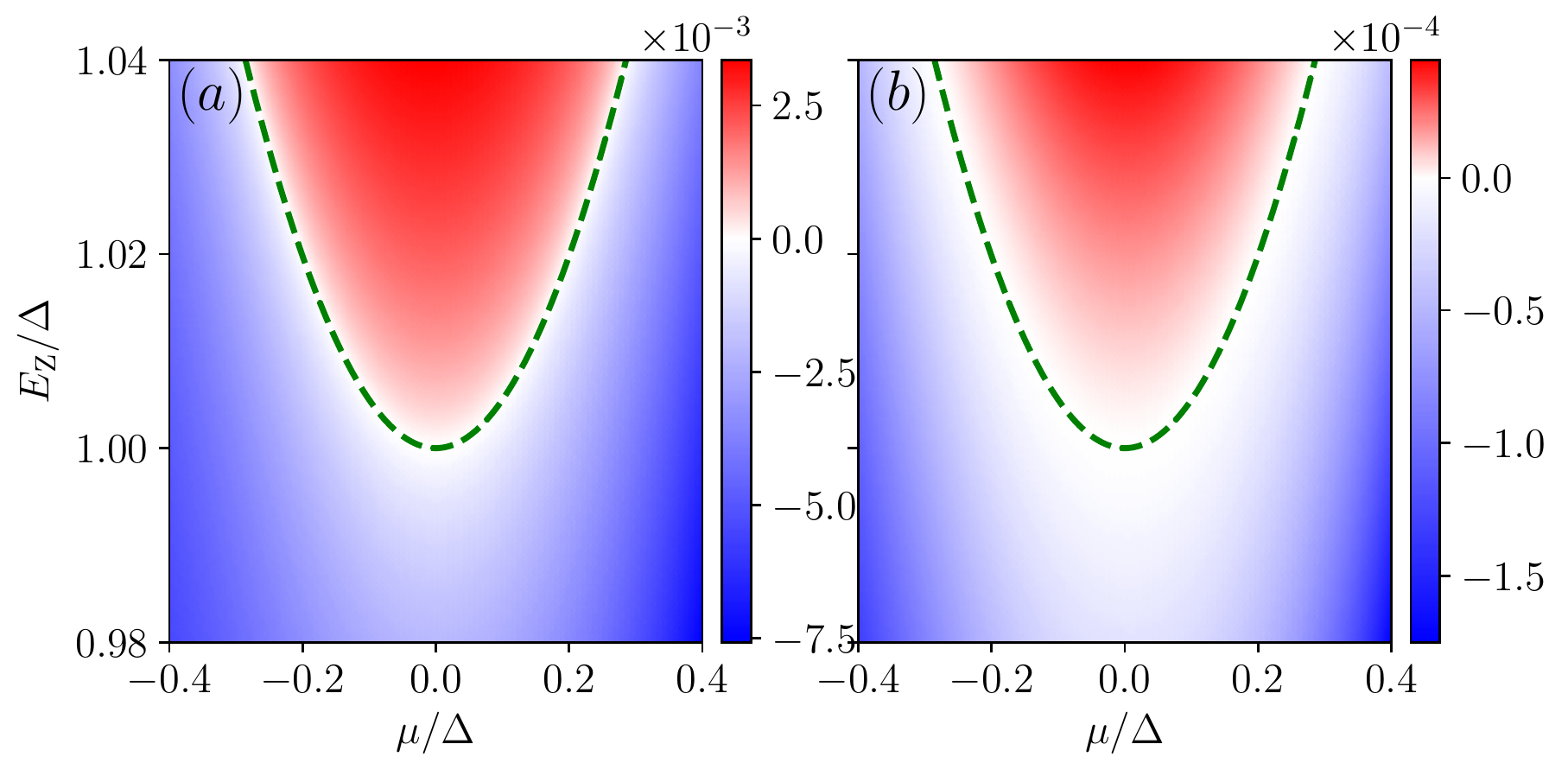}
\caption{Susceptibility $\chi''_d$, in units of $\Delta/\alpha^2$, in the (a) 1D model and, in units $\Delta^2/\alpha^3$, in the (b) 2D model as a function of chemical potential $\mu$ and Zeeman energy $E_{\rm Z}$. 
The response changes sign at the topological phase transition (green dashed line) $E_g=0$ or $E_{\rm Z}^2=\Delta^2+\mu^2$. 
The frequency is tuned at the resonance condition, $\hbar\omega=2|E_g|$.
}
\label{fig:topo_map}
\end{figure}

\section{The general response function}\label{sec:gen_sol}
In this section we present an analysis of the dynamical susceptibility valid for arbitrary driving frequency and choice of material parameters.
The particular limits, discussed before, are recovered from the more general expression presented here.

The response functions follows from Eq.~\eqref{chi_mat}. The full Matsubara Green's function is a $4\times4$ matrix that can be inverted analytically to give
\begin{equation}
G(\bm k,i\nu) = \frac{(i\nu+H)^2(i\nu-H)}{[(i\nu)^2 - E_+^2(k)][(i\nu)^2 - E_-^2(k)]},
\end{equation}
with the energies $E_\pm(k)$ from Eq.~\eqref{eig_energies}.
The susceptibility follows from Eq.~\eqref{chi_mat} after performing the trace over spin and particle-hole degrees of freedom and integrating over the Matsubara frequency.
The general result is quite lengthy, and it is relegated to Appendix~\ref{sec:response}. 
Nevertheless, it is further simplified near the transition by keeping in mind that the energy $E_+(k)$ is always much larger than the spectral gap [set by $E_-(k)$], namely $E_+(k)\gg E_-(k)$.
Considering dynamics on the scale of twice the gap $E_g$, allows us to neglect terms from high-frequency transitions, corresponding to $\hbar|\omega|>2E_+$ and $\hbar|\omega| > E_++E_-$. 

In the 2D model there is an additional angular integral which, due to the rotation symmetry of the Hamiltonian, is trivial and yields $2\pi$.
Therefore in both the 1D and 2D models, the response function reduces to a simple form involving a single integral over momenta.
After analytical continuation $i\omega\to \omega+i\delta/\hbar$, it reads
\begin{eqnarray}\label{gen_sol}
\chi_1(\omega) &=& \int \frac{dk}{2\pi} F(k),\quad
\chi_2(\omega) = \int_0^\infty \frac{dk k}{2\pi} F(k),\\
F(k)&=&
\frac{\alpha E_{\rm Z}\Delta^2k^2[E^2_{\rm Z}-\sqrt{\alpha^2k^2\xi^2+E_{\rm Z}^2(\Delta^2+\xi^2)}]}
{E_-^2(k)[\alpha^2k^2\xi^2+E_{\rm Z}^2(\Delta^2+\xi^2)]}\notag\\
&&\times\bigg[
\frac{1}{2E_-(k)-\hbar\omega-i\delta}
+
\frac{1}{2E_-(k)+\hbar\omega+i\delta}
\bigg].\notag
\end{eqnarray}
The remaining integral over momentum is performed numerically, usually with $\delta=0.004\Delta$.


The susceptibility recovers Eqs.~\eqref{chi_poles_1} and~\eqref{chi_poles_2}, which were obtained in the approximation $\xi\to 0$.
Therefore it recovers near the topological transition the square-root scaling with frequency for $\chi''_1(\omega)$ and, the linear one, for $\chi_2''(\omega)$.
A density plot for the susceptibility is shown in Fig.~\ref{fig:topo_map} in the $(\mu, E_{\rm Z})$ parameter space, with the frequencies tuned at the resonance condition.
As expected the dynamical susceptibility maps exactly the position of the topological phase transition $E_g=0$ and changes sign across it.
This confirms that the topological nontrivial phases could be identified by measuring $\chi''_d(\omega)$.

\section{Finite disordered systems}
\label{sec:finite_disorder_systems}
The present section studies the behavior of the Rashba susceptibility in finite-size tight-binding systems obtained in a lattice discretization of the continuum Hamiltonians for the 1D model Eq.~\eqref{1d_ham} and the 2D model Eq.~\eqref{2d_ham}.
The goal is to determine the robustness of our results when translation invariance is broken by disorder potentials.

The Rashba susceptibility in the finite system is defined from a Fourier transform of Eq.~\eqref{susc}:
\begin{equation}\label{chi_finite}
\chi_{jR} = \frac{1}{N}\sum_{mn}(f_m-f_n)\frac{\avg{m|\sigma_j|n}\avg{n|H_{\rm so}/\alpha|m}}
{\hbar\omega+E_m-E_n+i\delta},
\end{equation}
where $m$ and $n$ run over the eigenstates of the BdG Hamiltonian, and $N$ is the total number of sites in the system. The Fermi Dirac functions are $f_n=(e^{\beta E_n}+1)^{-1}$. 
It is implied that the Pauli matrix $\sigma_j$ acts as identity in the site space.

We consider Anderson disorder, such that the onsite chemical potential becomes a function of lattice sites, $\mu\to\mu_i=\mu+\Delta\mu_i$, at an arbitrary site $i$. 
The potentials $\Delta\mu_i$ are normally distributed random numbers with zero mean. 
The strength of disorder is parametrized in simulations by the standard deviation $\sigma$ of the normal distribution.
We explore two strong disorder strengths.
First for $\sigma=0.04\Delta$, the standard deviation is comparable to the spectral gaps $E_g$ which are typically considered in our setup. 
Secondly, the disorder is a factor of magnitude larger, $\sigma=0.4\Delta$, such that fluctuations in $\mu$ are comparable to the superconducting gap $\Delta$.
For comparison, we also plot the susceptibility in the clean infinite system, labeled by $\sigma=0$.

As in the previous sections, we denote and measure the non-vanishing Rasbha susceptibility $\chi_1\equiv\chi_{xR}$ in 1D and $\chi_2\equiv\chi_{zR}$, in 2D, for each disorder realization.
The results presented in the following are for the disorder-averaged susceptibility
\begin{equation}
\avg{\chi_{1,2}(\omega)} = \sum_{j=1}^M \chi^{(j)}_{1,2}(\omega),
\end{equation}
where $\chi^{(j)}$ stands for the susceptibility obtained for a specific disorder realization $j$. In simulations we usually average over $M=100$ disorder realizations at each different frequency.

\begin{figure}
	\includegraphics[width=\columnwidth]{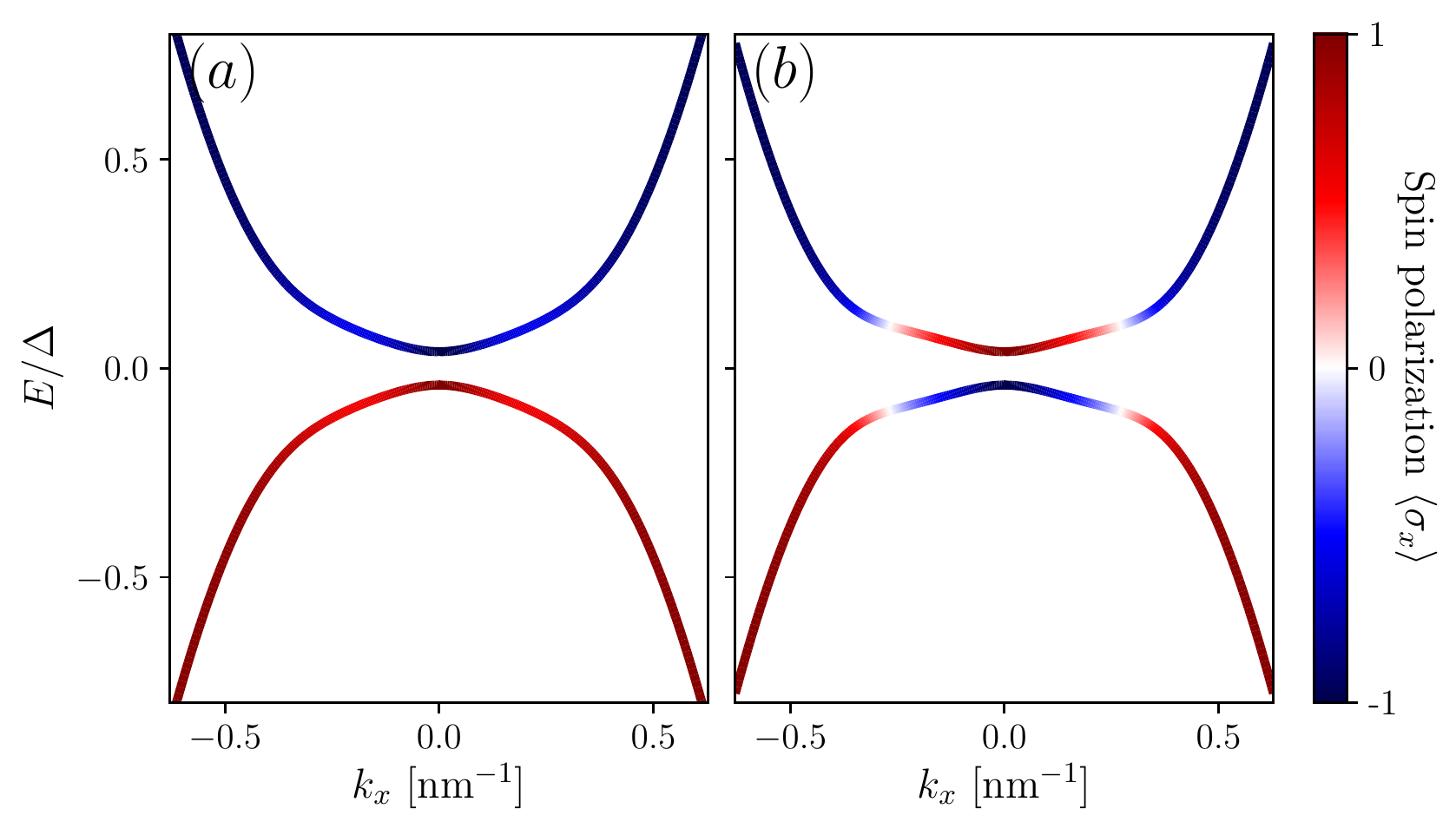}
	\caption{Spin polarization in the lowest energy bands of the 1D model~\eqref{1d_ham} for the modified system parameters used in the finite system simulations, $t=\SI{1}{meV}$, $\alpha = \SI{0.1}{nm\cdot meV}$, $\Delta=\SI{0.25}{meV}$, and $\mu=0$.
		The spectral gap is (a) $E_g=-0.04\Delta$, in the trivial phase, and (b) $E_g=0.04\Delta$, in the nontrivial phase.}
	\label{fig:spin_pol_mod}
\end{figure}

We perform exact diagonalization to obtain eigenenergies and eigenvectors in an energy window containing the spectral gap $E_g$.
Exact diagonalization imposes a certain limitation in exploring large systems. 
The mean level spacing varies as $t/N$, and when it become comparable to $E_g$, the simulations are no longer accurately describing the system at hand. 
Since increasing system size $N$ to the desired value is not always possible in simulations, the bandwidth is artificially reduced to $t=\SI{1}{meV}$ and the Rashba spin-orbit coupling strength, to $\alpha=\SI{0.1}{meV\cdot nm}$. 
This choice does not modify qualitatively the physics of the problem near the topological transition, and, in particular, the gap at $k=0$ remains the same.
An example of spin-polarization reversal at the topological transition is shown in Fig.~\ref{fig:spin_pol_mod}.

\subsection{1D model}
\begin{figure}[t]
	\includegraphics[width=\columnwidth]{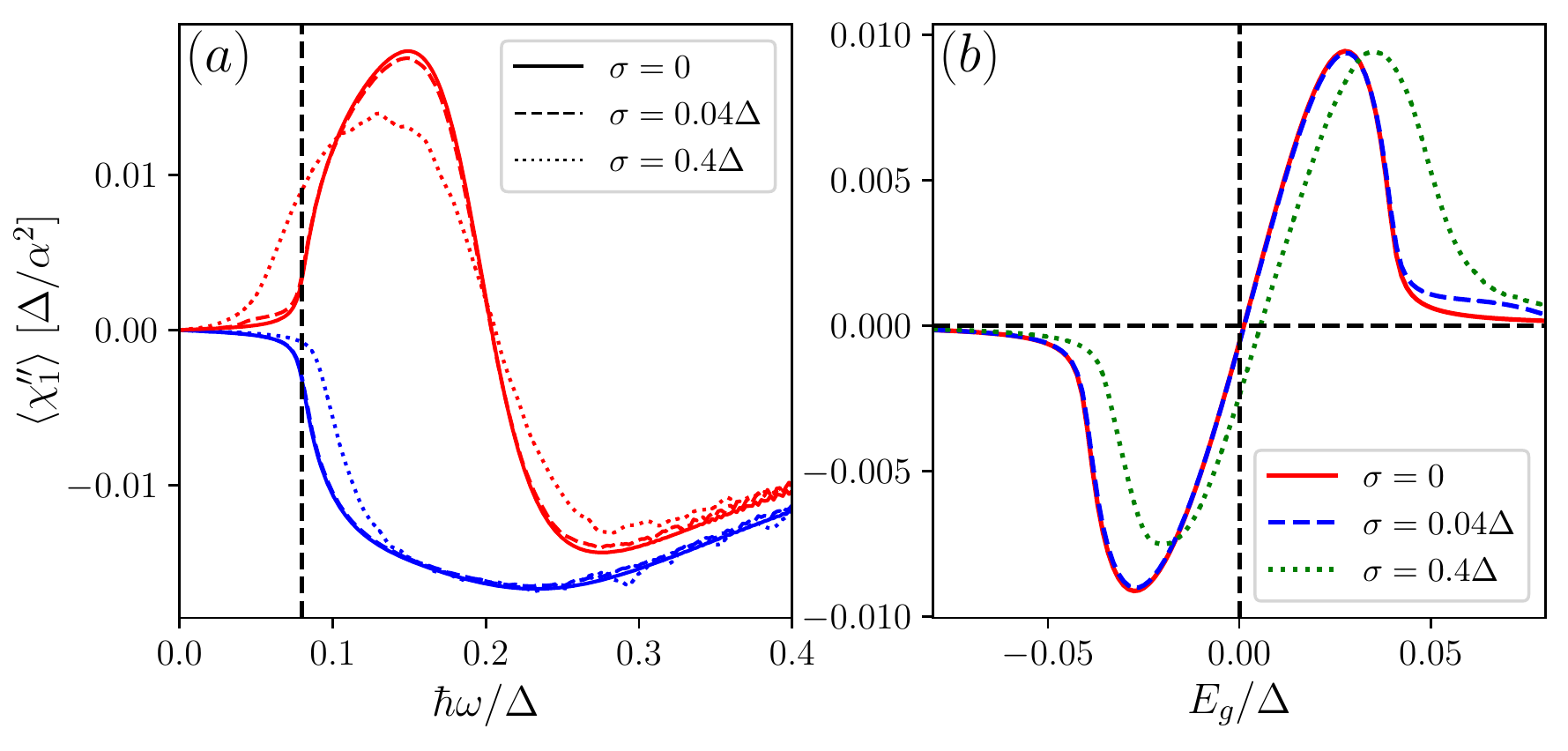}
	\caption{Imaginary part of the disorder-averaged Rashba susceptibility in the discretized 1D nanowire Eq.~\eqref{1d_ham} on 1200 sites, and for 100 disorder realizations at each frequency.
		The lines corresponding to $\sigma=0$ are obtained in the infinite clean system.
		(a) Rashba susceptibility as a function of frequency in the topologically nontrivial phase $E_g=0.04\Delta$ (red lines) and trivial phase $E_g=-0.04\Delta$ (blue lines). The dashed vertical line at $2|E_g|$ marks the excitation threshold in the clean system.
		(b) Rashba susceptibility near the transition, as a function of the gap $E_g$ at fixed frequency $\hbar\omega=0.08\Delta$.
	}
	\label{fig:dis_1d}
\end{figure}

In 1D we consider tight-binding systems with sizes $N\sim 1000$ sites and compute the disorder-averaged Rashba susceptibility either as a function of frequency, for a fixed absolute value of the gap $E_g$ in Fig.~\ref{fig:dis_1d}(a) (in both trivial and nontrivial phases), or at fixed frequency, but varying the spectral gap over the topological transition in Fig.~\ref{fig:dis_1d}(b).
[In the latter case, for each disorder realization, we obtain a susceptibility curve as a function of $E_g$ (or Zeeman energy)].
We see that the results are quite robust even when the fluctuations in the chemical potential have a standard deviation comparable to the gap $E_g$, and the results results for $\sigma=0.04\Delta$ follow closely the results in the infinite clean system.
In Fig.~\ref{fig:dis_1d}(a) the susceptibility has different signs in the trivial and nontrivial phase and near the excitation threshold of $2|E_g|$ has the predicted squared-root behavior as a function of frequency.
A noticeable difference with previous results in Fig.~\ref{fig:img_chi_mu0} is the signal changing sign at $\hbar\omega\simeq 0.2\Delta$ in the nontrivial phase.
This phenomenon is due to the spin polarization change in the nontrivial phase present already for an infinite clean system, occurring for the modified system parameters used in simulations at $\hbar\omega\simeq 0.1\Delta$ [see Fig.~\ref{fig:spin_pol_mod}(b)].
When disorder is comparable with the superconducting gap, more states enter in the energy window defined by the clean system gap $E_g$, which leads to noticeable differences from the clean system.
In particular, Fig.~\ref{fig:dis_1d}(b) shows a Rashba susceptibility curve for disorder strength $\sigma=0.4\Delta$ that is slightly shifted and not changing its sign exactly at the topological transition.
Therefore, in the 1D case, we conclude that the theoretical predictions from the clean model can be trusted when potential disorder strength is comparable to the spectral gap $E_g$.

\subsection{2D model}
In 2D we consider a tight-binding Hamiltonian obtained from a discretization of Eq.~\eqref{2d_ham} on a square lattice of size $120\times 120$ sites and record the mean Rashba susceptibility, averaged over 100 disorder realizations at each frequency. 
When the clean system enters a topologically nontrivial phase, Majorana edge states form at the perimeter of the patch and populate the low-energy space.
To gain better resolution for the bulk response of the system, we eliminate the edges by imposing periodic boundary conditions to the disordered system.
The tight-binding modeling of the finite 2D system is done using the \textsc{Kwant} package~\cite{Groth2014}.
For the rest, we perform the same numerical experiments as in the 1D case.

\begin{figure}[t]
	\includegraphics[width=\columnwidth]{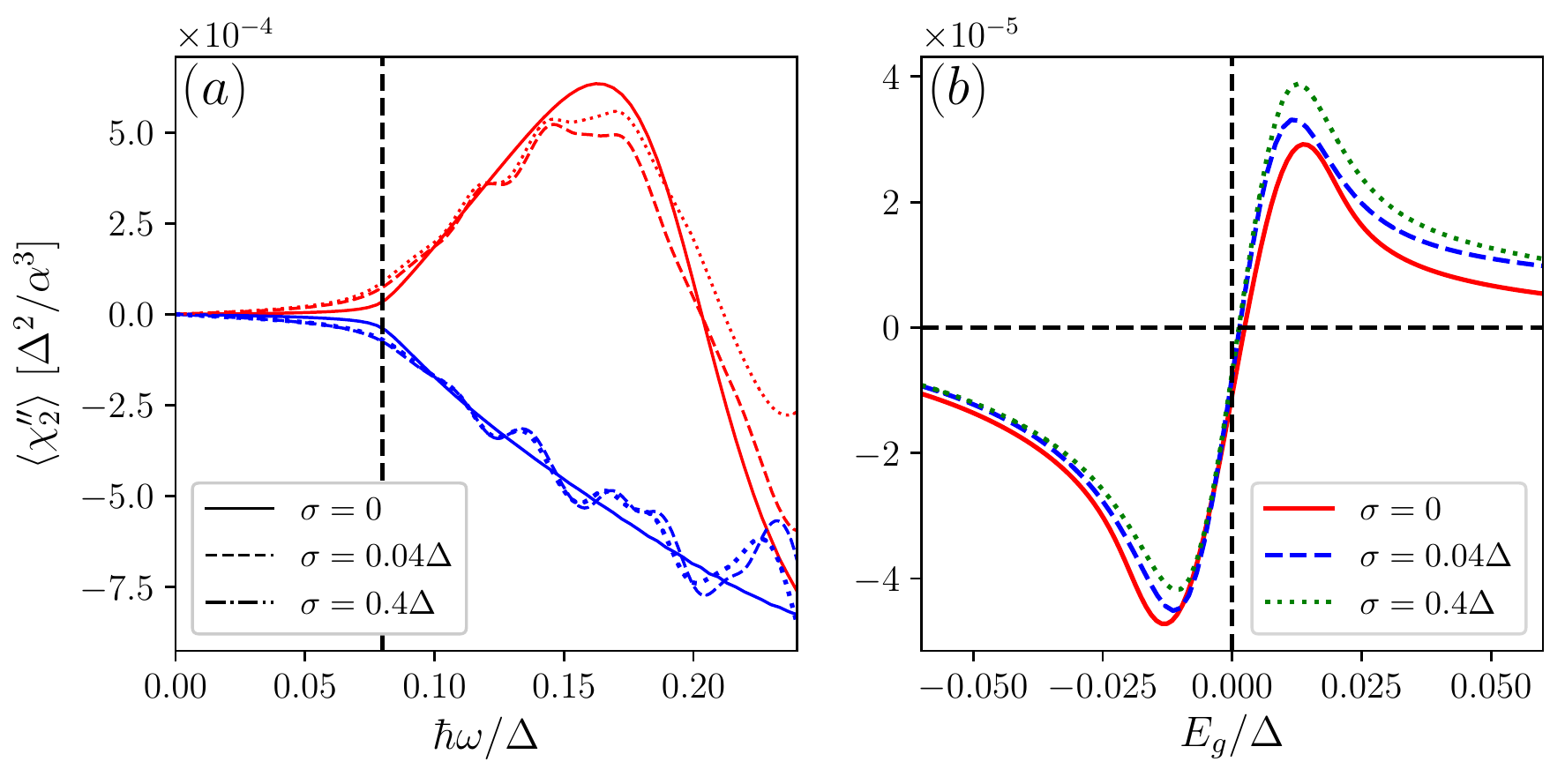}
	\caption{Imaginary part of the disordered-averaged Rashba susceptibility for a tight-binding model which discretizes Eq.~\eqref{2d_ham} on a $120\times 120$ site periodic patch for two disorder strengths, with 100 disorder realizations at each frequency, and in an infinite system ($\sigma=0$). 
		(a) Susceptibility as a function of frequency in the topologically nontrivial phase ($E_g=0.04\Delta$, red curves) and trivial phase ($E_g=-0.04\Delta$, blue curves).
		The vertical line marks the excitation threshold $2|E_g|$. (b) Susceptibility as a function of the spectral gap at fixed driving frequency $\hbar\omega = 0.04\Delta$.
	}
	\label{fig:dis_2d}
\end{figure}

Numerical results in the 2D model are displayed in Fig.~\ref{fig:dis_2d}, and show that the predictions from the clean system are even more robust to disorder compared to the 1D case. 
When disorder strength is either comparable to the spectral gaps, or with the superconducting gap $\Delta$, the susceptibility curve follows the clean system results and shows the expected linear scaling with frequency near the excitation threshold [see Fig.~\ref{fig:dis_2d}(a)].
Large oscillations in the curves developing at higher frequency are due to finite-size effects which we could not eliminate completely for the chosen system sizes even by increasing the broadening to $\delta=0.012\Delta$.
The larger $\delta$ is also responsible for an overall slight shift in susceptibility curves as a function of the spectral gap in Fig.~\ref{fig:dis_2d}(b) and a nonzero susceptibility below $2|E_g|$ in Fig.~\ref{fig:dis_2d}(a).

Therefore the expected change in the system response recorded in the Rashba susceptibility remains quite robust to Anderson disorder both in one and two dimensions for disorder strength comparable to the spectral gap $E_g$.

\section{Conclusion}
\label{sec:conc}
In this paper, we studied theoretically two model systems of semiconductor-superconductor heterostructures which support Majorana bound states.
We proposed an all-electric spectroscopic method to discriminate the topological phases in such materials by exploiting the bulk spin inversion at the topological transition.
Our proposal uses time-varying electric fields, which dynamically modulate the Rashba spin-orbit coupling strength of the semiconductor, and cause resonant transitions between the electronic bands.
Relaxation processes are then measured by optical spectroscopy at microwave frequencies using, for example, techniques developed in electron spin resonance spectroscopy.
The above protocol is modeled within linear response theory by a modified susceptibility.
We showed that its imaginary part, $\chi''(\omega)$, can be used to discriminate the topological phases, since spin relaxation processes depend on the sign of the spectral gap close to the transition.
The response is robust to potential disorder in the system and could be distinguished from the trivial response due to the electric field coupling to the electronic density. 
Such measurements may therefore be used to detect the topological nontrivial phases without the need to access information about the localized Majorana modes hosted in them.

\acknowledgments{
We thank P.~Szumniak for drawing our attention to Ref.~\onlinecite{Szumniak2017} on the subject of bulk spin polarization in the Rashba nanowire-superconductor heterostructures and to M.P.~Nowak for a critical reading of the manuscript.
D.~S.~was supported by the The Romanian National Authority for Scientific Research and Innovation, CNCS-UEFISCDI, through the Project No.~PN-III-P1-1.2-PCCDI-2017-0338, and, additionally, for a mobility, through the Project No.~32PFE/19.10.2018.
B.~D.~acknowledges support from the National Research, Development and Innovation Office---NKFIH  within the Quantum Technology National Excellence Program (Project No.~2017-1.2.1-NKP-2017-00001), K119442.}

\appendix
\section{Response due to time-varying chemical potential}\label{sec:dens_resp}
The alternating electric field applied to the system couples to the electronic density in the system, leading to fluctuations in the chemical potential. 
This section studies how the chemical potential variation affects the spin polarization.

The change in polarization due to changes in the chemical potential is given in linear response theory by
\begin{equation}
\avg{\delta \sigma_j(t)} = \int_{-\infty}^t dt'\chi_{j\mu}(t-t')\delta\mu(t'),
\end{equation}
with susceptibility
\begin{equation}
\chi_{j\mu}(t-t')=-i\theta(t-t')\avg{[\sigma_j(t),-\tau_z(t')]}.
\end{equation}
Assuming time-translation and spatial-translation invariance, we compute the susceptibility $\chi_{j\mu}$ using Matsubara Green's functions $G$ of the system in absence of the perturbation, in the zero-temperature limit,
\begin{equation}
\chi_{j\mu}(i\omega) = -\int\frac{d\nu d^d\bm k}{(2\pi)^{d+1}} 
\text{Tr}[\sigma_j G(\bm k, i\nu) \tau_z G(\bm k, i\omega+i\nu)].
\end{equation}
We determine in the 1D model $\chi_{1\mu}\equiv\chi_{x\mu}$ and in the 2D one, $\chi_{2\mu}\equiv\chi_{z\mu}$, after performing the integration over Matsubara frequencies $\nu$ and analytical continuation $i\omega \to \omega+i\delta/\hbar$. 
In the low-frequency approximation, where only transitions between the lowest quasiparticle bands are allowed, the susceptibility reads
\begin{eqnarray}
\chi_{1\mu}(\omega)&=& \int_{-\infty}^\infty \frac{dk}{2\pi} F_{\mu},\quad
\chi_{2\mu}(\omega)= \int_{0}^\infty \frac{dk}{2\pi} F_{\mu},\notag\\
\notag\\
F_\mu(k,\omega)&=& -\frac{E_{\rm Z}\alpha^2k^2\Delta^2\xi}
{E_-^2(k)[\alpha^2k^2\xi^2+E_{\rm Z}^2(\Delta^2+\xi^2)]}\\
&&\times\bigg[
\frac{1}{2E_-(k)-\hbar\omega-i\delta}
+
\frac{1}{2E_-(k)+\hbar\omega+i\delta}
\bigg].\notag
\end{eqnarray}
The imaginary part of the susceptibility, is further obtained by numerical integration over momenta $k=k_x$ in 1D and $k=(k_x^2+k_y^2)^{1/2}$ in 2D.
Several properties can be immediately identified.

\begin{figure}[t]
\includegraphics[width=\columnwidth]{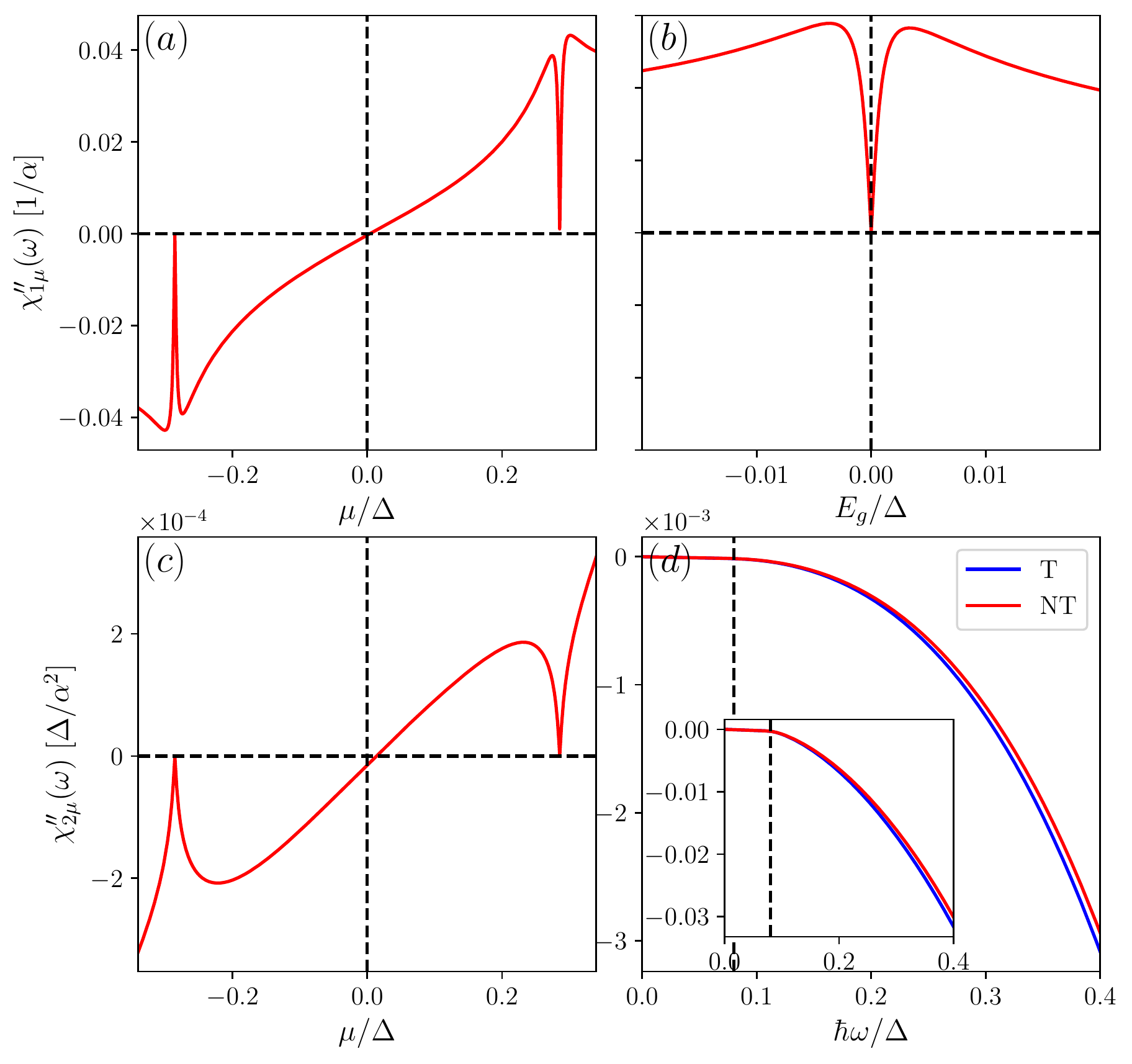}
\caption{Imaginary susceptibility $\chi_{1\mu}''$ in 1D, in units $1/\alpha$, and $\chi_{2\mu}''$ in 2D, in units $\Delta/\alpha^2$.
(a, c) Susceptibility $\chi_{1\mu}''$, respectively $\chi_{2\mu}''$, at the resonance condition $\hbar\omega=2|E_g|$, when the chemical potential $\mu$ varies and $E_{\rm Z}=1.04\Delta$. At finite $|\mu|$, a dip in the response marks the topological transition. 
(b) A zoom for the 1D case at a topological transition for a fixed $\mu=0.4\Delta$. The system crosses the topological transition when varying the gap $E_g$, by changing the Zeeman field strength.
(d) Susceptibility $\chi''_2$ as a function of frequency behaves similarly in the topologically nontrivial phase (blue line, NT, at $E_g=0.04\Delta$) and in the trivial phase (red line, T, at $E_g=-0.04\Delta$) at $\mu=0$. Similar results are presented in the inset for $\chi''_1$.
In contrast to the Rashba susceptibility there is no sign change in response at the topological transition.}
\label{fig:dens_resp}
\end{figure}

Near the topological transition at $k=0$ the contribution to the imaginary susceptibility of terms $\mathcal O(k^2)$ in the dispersion $\xi$ is small. 
This leads to $\chi_{j\mu}\propto \mu$. This shows that it is expected to have an imaginary response which changes sign at $\mu=0$ and it is linear in $\mu$. This is qualitatively correct, as shown in numerical results in Fig.~\ref{fig:dens_resp}(a, c).
When the frequency is tuned at the resonance condition, and the system is at the topological transition, $\chi_{j\mu}$ falls to zero due to the momentum dependence in $F_\mu$ Fig.~\ref{fig:dens_resp}(a-c). The response $\chi''_{j\mu}$ close to the transition is even in the spectral gap $E_g$ [see Fig.~\ref{fig:dens_resp}(b)], in contrast to the Rashba susceptibility, which is odd in $E_g$.
Therefore in an experiment for different values of $E_g$, it would be possible to discriminate between the linear response due to chemical potential modulation from the Rashba susceptibility by taking linear combinations of the signal, for opposite values of the gap $E_g$.

The trivial behavior of the response due to modulation of the chemical potential is seen also as a function of frequency in Fig.~\ref{fig:dens_resp}(d), where the response in either of the topological phases is almost identical, for the same magnitude of the gap $|E_g|$. 
Moreover, the threshold for excitations $2|E_g|$ is not changed compared to the Rasbha susceptibility. This also indicates that the time-varying chemical potential does not modify the spectral gap in the system at the linear response level.

\section{The high-frequency response function}
\label{sec:response}
Eq.~\eqref{gen_sol} contains the general \textit{low-frequency} Rashba response function for both 1D and 2D models, for arbitrary choice of material parameters.
Here we present the susceptibility containing the information about transitions between all quasiparticle bands, at arbitrary frequency. 
After tracing out the particle-hole and spin degrees of freedom and performing the integral over the Matsubara frequencies in Eq.~\eqref{chi_mat}, it reads
\begin{widetext}
\begin{eqnarray}
\chi_1(i\omega) &=& \int \frac{dk}{2\pi} \frac{N(k)}{D(k)},\quad \chi_2(i\omega) = \int_0^{\infty}\frac{dk k}{2\pi} \frac{N(k)}{D(k)},\notag\\
N(k)&=&8 \alpha E_{\rm Z} k^2\Big\{ E_-^3 E_+(4E_+^2+\hbar^2\omega^2)
+E_-^2
\{4[E_{\rm Z}^4+\alpha^4 k^4-2\Delta^2\alpha^2 k^2-3\Delta^4-2(\alpha^2 k^2+\Delta^2)\xi^2+\xi^4\notag\\
&&+\,2E_+^2(E_{\rm Z}^2+\alpha^2k^2-\Delta^2-3\xi^2)+2E_{\rm Z}^2(\alpha^2 k^2+\Delta^2-\xi^2)]
-[E_+^2+4(\Delta^2+\xi^2-\alpha^2k^2-E_{\rm Z}^2)]\hbar^2\omega^2\}\notag\\
&&+\,E_-E_+\{12[E_{\rm Z}^4+\alpha^4k^4-3\Delta^4-2\Delta^2\xi^2+\xi^4+2E_{\rm Z}^2(\alpha^2k^2+\Delta^2-\xi^2)
-2\alpha^2k^2(\Delta^2+\xi^2)]\notag\\
&&+\, [E_+^2+6(E_{\rm Z}^2+\alpha^2k^2-\Delta^2+\xi^2)]\hbar^2\omega^2+\hbar^4\omega^4\}\notag\\
&&+\,(4E_+^2+\hbar^2\omega^2)\{E_{\rm Z}^4+\alpha^4k^4+E_{\rm Z}^2(2\alpha^2k^2+2\Delta^2-2\xi^2+\hbar^2\omega^2)
-(\Delta^2+\xi^2)(3\Delta^2-\xi^2+\hbar^2\omega^2)\notag\\
&&+\,\alpha^2k^2[\hbar^2\omega^2-2(\Delta^2+\xi^2)]\}
\Big\}\notag\\
D(k)&=&E_-E_+(E_-+E_+)(4E_-^2+\hbar^2\omega^2)(4E_+^2+\hbar^2\omega^2)[(E_-+E_+)^2+\hbar^2\omega^2],
\end{eqnarray}
with $k$ changing accordingly in the 1D and 2D system.
\end{widetext}
After analytical continuation of $\omega$,
we have checked that the above expression gives the same results as those of Eq.~\eqref{susc}, where the Brillouin zone is discretized, Hamiltonian eigenstates are obtained at each momentum, and all integrals are carried out numerically.

\bibliographystyle{apsrev4-2}
\bibliography{bibl}

\begin{thebibliography}{72}%
\makeatletter
\providecommand \@ifxundefined [1]{%
 \@ifx{#1\undefined}
}%
\providecommand \@ifnum [1]{%
 \ifnum #1\expandafter \@firstoftwo
 \else \expandafter \@secondoftwo
 \fi
}%
\providecommand \@ifx [1]{%
 \ifx #1\expandafter \@firstoftwo
 \else \expandafter \@secondoftwo
 \fi
}%
\providecommand \natexlab [1]{#1}%
\providecommand \enquote  [1]{``#1''}%
\providecommand \bibnamefont  [1]{#1}%
\providecommand \bibfnamefont [1]{#1}%
\providecommand \citenamefont [1]{#1}%
\providecommand \href@noop [0]{\@secondoftwo}%
\providecommand \href [0]{\begingroup \@sanitize@url \@href}%
\providecommand \@href[1]{\@@startlink{#1}\@@href}%
\providecommand \@@href[1]{\endgroup#1\@@endlink}%
\providecommand \@sanitize@url [0]{\catcode `\\12\catcode `\$12\catcode
  `\&12\catcode `\#12\catcode `\^12\catcode `\_12\catcode `\%12\relax}%
\providecommand \@@startlink[1]{}%
\providecommand \@@endlink[0]{}%
\providecommand \url  [0]{\begingroup\@sanitize@url \@url }%
\providecommand \@url [1]{\endgroup\@href {#1}{\urlprefix }}%
\providecommand \urlprefix  [0]{URL }%
\providecommand \Eprint [0]{\href }%
\providecommand \doibase [0]{https://doi.org/}%
\providecommand \selectlanguage [0]{\@gobble}%
\providecommand \bibinfo  [0]{\@secondoftwo}%
\providecommand \bibfield  [0]{\@secondoftwo}%
\providecommand \translation [1]{[#1]}%
\providecommand \BibitemOpen [0]{}%
\providecommand \bibitemStop [0]{}%
\providecommand \bibitemNoStop [0]{.\EOS\space}%
\providecommand \EOS [0]{\spacefactor3000\relax}%
\providecommand \BibitemShut  [1]{\csname bibitem#1\endcsname}%
\let\auto@bib@innerbib\@empty
\bibitem [{\citenamefont {Kitaev}(2001)}]{Kitaev2001}%
  \BibitemOpen
  \bibfield  {author} {\bibinfo {author} {\bibfnamefont {A.~Y.}\ \bibnamefont
  {Kitaev}},\ }\href {https://doi.org/10.1070/1063-7869/44/10S/S29} {\bibfield
  {journal} {\bibinfo  {journal} {Phys. Usp.}\ }\textbf {\bibinfo {volume}
  {44}},\ \bibinfo {pages} {131} (\bibinfo {year} {2001})}\BibitemShut
  {NoStop}%
\bibitem [{\citenamefont {Nayak}\ \emph {et~al.}(2008)\citenamefont {Nayak},
  \citenamefont {Simon}, \citenamefont {Stern}, \citenamefont {Freedman},\ and\
  \citenamefont {Das~Sarma}}]{Nayak2008}%
  \BibitemOpen
  \bibfield  {author} {\bibinfo {author} {\bibfnamefont {C.}~\bibnamefont
  {Nayak}}, \bibinfo {author} {\bibfnamefont {S.~H.}\ \bibnamefont {Simon}},
  \bibinfo {author} {\bibfnamefont {A.}~\bibnamefont {Stern}}, \bibinfo
  {author} {\bibfnamefont {M.}~\bibnamefont {Freedman}},\ and\ \bibinfo
  {author} {\bibfnamefont {S.}~\bibnamefont {Das~Sarma}},\ }\href
  {https://doi.org/10.1103/RevModPhys.80.1083} {\bibfield  {journal} {\bibinfo
  {journal} {Rev. Mod. Phys.}\ }\textbf {\bibinfo {volume} {80}},\ \bibinfo
  {pages} {1083} (\bibinfo {year} {2008})}\BibitemShut {NoStop}%
\bibitem [{\citenamefont {Fu}\ and\ \citenamefont {Kane}(2008)}]{Fu2008}%
  \BibitemOpen
  \bibfield  {author} {\bibinfo {author} {\bibfnamefont {L.}~\bibnamefont
  {Fu}}\ and\ \bibinfo {author} {\bibfnamefont {C.~L.}\ \bibnamefont {Kane}},\
  }\href {https://doi.org/10.1103/PhysRevLett.100.096407} {\bibfield  {journal}
  {\bibinfo  {journal} {Phys. Rev. Lett.}\ }\textbf {\bibinfo {volume} {100}},\
  \bibinfo {pages} {096407} (\bibinfo {year} {2008})}\BibitemShut {NoStop}%
\bibitem [{\citenamefont {Oreg}\ \emph {et~al.}(2010)\citenamefont {Oreg},
  \citenamefont {Refael},\ and\ \citenamefont {von Oppen}}]{Oreg2010}%
  \BibitemOpen
  \bibfield  {author} {\bibinfo {author} {\bibfnamefont {Y.}~\bibnamefont
  {Oreg}}, \bibinfo {author} {\bibfnamefont {G.}~\bibnamefont {Refael}},\ and\
  \bibinfo {author} {\bibfnamefont {F.}~\bibnamefont {von Oppen}},\ }\href
  {https://doi.org/10.1103/PhysRevLett.105.177002} {\bibfield  {journal}
  {\bibinfo  {journal} {Phys. Rev. Lett.}\ }\textbf {\bibinfo {volume} {105}},\
  \bibinfo {pages} {177002} (\bibinfo {year} {2010})}\BibitemShut {NoStop}%
\bibitem [{\citenamefont {Lutchyn}\ \emph {et~al.}(2010)\citenamefont
  {Lutchyn}, \citenamefont {Sau},\ and\ \citenamefont
  {Das~Sarma}}]{Lutchyn2010}%
  \BibitemOpen
  \bibfield  {author} {\bibinfo {author} {\bibfnamefont {R.~M.}\ \bibnamefont
  {Lutchyn}}, \bibinfo {author} {\bibfnamefont {J.~D.}\ \bibnamefont {Sau}},\
  and\ \bibinfo {author} {\bibfnamefont {S.}~\bibnamefont {Das~Sarma}},\ }\href
  {https://doi.org/10.1103/PhysRevLett.105.077001} {\bibfield  {journal}
  {\bibinfo  {journal} {Phys. Rev. Lett.}\ }\textbf {\bibinfo {volume} {105}},\
  \bibinfo {pages} {077001} (\bibinfo {year} {2010})}\BibitemShut {NoStop}%
\bibitem [{\citenamefont {Sau}\ \emph {et~al.}(2010)\citenamefont {Sau},
  \citenamefont {Lutchyn}, \citenamefont {Tewari},\ and\ \citenamefont
  {Das~Sarma}}]{Sau2010}%
  \BibitemOpen
  \bibfield  {author} {\bibinfo {author} {\bibfnamefont {J.~D.}\ \bibnamefont
  {Sau}}, \bibinfo {author} {\bibfnamefont {R.~M.}\ \bibnamefont {Lutchyn}},
  \bibinfo {author} {\bibfnamefont {S.}~\bibnamefont {Tewari}},\ and\ \bibinfo
  {author} {\bibfnamefont {S.}~\bibnamefont {Das~Sarma}},\ }\href
  {https://doi.org/10.1103/PhysRevLett.104.040502} {\bibfield  {journal}
  {\bibinfo  {journal} {Phys. Rev. Lett.}\ }\textbf {\bibinfo {volume} {104}},\
  \bibinfo {pages} {040502} (\bibinfo {year} {2010})}\BibitemShut {NoStop}%
\bibitem [{\citenamefont {Choy}\ \emph {et~al.}(2011)\citenamefont {Choy},
  \citenamefont {Edge}, \citenamefont {Akhmerov},\ and\ \citenamefont
  {Beenakker}}]{Choy2011}%
  \BibitemOpen
  \bibfield  {author} {\bibinfo {author} {\bibfnamefont {T.-P.}\ \bibnamefont
  {Choy}}, \bibinfo {author} {\bibfnamefont {J.~M.}\ \bibnamefont {Edge}},
  \bibinfo {author} {\bibfnamefont {A.~R.}\ \bibnamefont {Akhmerov}},\ and\
  \bibinfo {author} {\bibfnamefont {C.~W.~J.}\ \bibnamefont {Beenakker}},\
  }\href {https://doi.org/10.1103/PhysRevB.84.195442} {\bibfield  {journal}
  {\bibinfo  {journal} {Phys. Rev. B}\ }\textbf {\bibinfo {volume} {84}},\
  \bibinfo {pages} {195442} (\bibinfo {year} {2011})}\BibitemShut {NoStop}%
\bibitem [{\citenamefont {Nadj-Perge}\ \emph {et~al.}(2013)\citenamefont
  {Nadj-Perge}, \citenamefont {Drozdov}, \citenamefont {Bernevig},\ and\
  \citenamefont {Yazdani}}]{Nadj-Perge2013}%
  \BibitemOpen
  \bibfield  {author} {\bibinfo {author} {\bibfnamefont {S.}~\bibnamefont
  {Nadj-Perge}}, \bibinfo {author} {\bibfnamefont {I.~K.}\ \bibnamefont
  {Drozdov}}, \bibinfo {author} {\bibfnamefont {B.~A.}\ \bibnamefont
  {Bernevig}},\ and\ \bibinfo {author} {\bibfnamefont {A.}~\bibnamefont
  {Yazdani}},\ }\href {https://doi.org/10.1103/PhysRevB.88.020407} {\bibfield
  {journal} {\bibinfo  {journal} {Phys. Rev. B}\ }\textbf {\bibinfo {volume}
  {88}},\ \bibinfo {pages} {020407} (\bibinfo {year} {2013})}\BibitemShut
  {NoStop}%
\bibitem [{\citenamefont {Braunecker}\ and\ \citenamefont
  {Simon}(2013)}]{Braunecker2013}%
  \BibitemOpen
  \bibfield  {author} {\bibinfo {author} {\bibfnamefont {B.}~\bibnamefont
  {Braunecker}}\ and\ \bibinfo {author} {\bibfnamefont {P.}~\bibnamefont
  {Simon}},\ }\href {https://doi.org/10.1103/PhysRevLett.111.147202} {\bibfield
   {journal} {\bibinfo  {journal} {Phys. Rev. Lett.}\ }\textbf {\bibinfo
  {volume} {111}},\ \bibinfo {pages} {147202} (\bibinfo {year}
  {2013})}\BibitemShut {NoStop}%
\bibitem [{\citenamefont {Klinovaja}\ \emph {et~al.}(2013)\citenamefont
  {Klinovaja}, \citenamefont {Stano}, \citenamefont {Yazdani},\ and\
  \citenamefont {Loss}}]{Klinovaja2013}%
  \BibitemOpen
  \bibfield  {author} {\bibinfo {author} {\bibfnamefont {J.}~\bibnamefont
  {Klinovaja}}, \bibinfo {author} {\bibfnamefont {P.}~\bibnamefont {Stano}},
  \bibinfo {author} {\bibfnamefont {A.}~\bibnamefont {Yazdani}},\ and\ \bibinfo
  {author} {\bibfnamefont {D.}~\bibnamefont {Loss}},\ }\href
  {https://doi.org/10.1103/PhysRevLett.111.186805} {\bibfield  {journal}
  {\bibinfo  {journal} {Phys. Rev. Lett.}\ }\textbf {\bibinfo {volume} {111}},\
  \bibinfo {pages} {186805} (\bibinfo {year} {2013})}\BibitemShut {NoStop}%
\bibitem [{\citenamefont {Alicea}(2012)}]{Alicea2012}%
  \BibitemOpen
  \bibfield  {author} {\bibinfo {author} {\bibfnamefont {J.}~\bibnamefont
  {Alicea}},\ }\href {https://doi.org/10.1088/0034-4885/75/7/076501} {\bibfield
   {journal} {\bibinfo  {journal} {Rep. Prog. Phys.}\ }\textbf {\bibinfo
  {volume} {75}},\ \bibinfo {pages} {076501} (\bibinfo {year}
  {2012})}\BibitemShut {NoStop}%
\bibitem [{\citenamefont {Leijnse}\ and\ \citenamefont
  {Flensberg}(2012)}]{Leijnse2012}%
  \BibitemOpen
  \bibfield  {author} {\bibinfo {author} {\bibfnamefont {M.}~\bibnamefont
  {Leijnse}}\ and\ \bibinfo {author} {\bibfnamefont {K.}~\bibnamefont
  {Flensberg}},\ }\href {https://doi.org/10.1088/0268-1242/27/12/124003}
  {\bibfield  {journal} {\bibinfo  {journal} {Semicond. Sci. Technol.}\
  }\textbf {\bibinfo {volume} {27}},\ \bibinfo {pages} {124003} (\bibinfo
  {year} {2012})}\BibitemShut {NoStop}%
\bibitem [{\citenamefont {Beenakker}(2013)}]{Beenakker2013}%
  \BibitemOpen
  \bibfield  {author} {\bibinfo {author} {\bibfnamefont {C.}~\bibnamefont
  {Beenakker}},\ }\href
  {https://doi.org/10.1146/annurev-conmatphys-030212-184337} {\bibfield
  {journal} {\bibinfo  {journal} {Annu. Rev. Condens. Matter Phys.}\ }\textbf
  {\bibinfo {volume} {4}},\ \bibinfo {pages} {113} (\bibinfo {year}
  {2013})}\BibitemShut {NoStop}%
\bibitem [{\citenamefont {Sarma}\ \emph {et~al.}(2015)\citenamefont {Sarma},
  \citenamefont {Freedman},\ and\ \citenamefont {Nayak}}]{Sarma2015}%
  \BibitemOpen
  \bibfield  {author} {\bibinfo {author} {\bibfnamefont {S.~D.}\ \bibnamefont
  {Sarma}}, \bibinfo {author} {\bibfnamefont {M.}~\bibnamefont {Freedman}},\
  and\ \bibinfo {author} {\bibfnamefont {C.}~\bibnamefont {Nayak}},\ }\href
  {https://doi.org/10.1038/npjqi.2015.1} {\bibfield  {journal} {\bibinfo
  {journal} {npj Quantum Inf.}\ }\textbf {\bibinfo {volume} {1}} (\bibinfo
  {year} {2015})}\BibitemShut {NoStop}%
\bibitem [{\citenamefont {Elliott}\ and\ \citenamefont
  {Franz}(2015)}]{Elliott2015}%
  \BibitemOpen
  \bibfield  {author} {\bibinfo {author} {\bibfnamefont {S.~R.}\ \bibnamefont
  {Elliott}}\ and\ \bibinfo {author} {\bibfnamefont {M.}~\bibnamefont
  {Franz}},\ }\href {https://doi.org/10.1103/RevModPhys.87.137} {\bibfield
  {journal} {\bibinfo  {journal} {Rev. Mod. Phys.}\ }\textbf {\bibinfo {volume}
  {87}},\ \bibinfo {pages} {137} (\bibinfo {year} {2015})}\BibitemShut
  {NoStop}%
\bibitem [{\citenamefont {Sato}\ and\ \citenamefont
  {Fujimoto}(2016)}]{Sato2016}%
  \BibitemOpen
  \bibfield  {author} {\bibinfo {author} {\bibfnamefont {M.}~\bibnamefont
  {Sato}}\ and\ \bibinfo {author} {\bibfnamefont {S.}~\bibnamefont
  {Fujimoto}},\ }\href {https://doi.org/10.7566/jpsj.85.072001} {\bibfield
  {journal} {\bibinfo  {journal} {J. Phys. Soc. Jpn.}\ }\textbf {\bibinfo
  {volume} {85}},\ \bibinfo {pages} {072001} (\bibinfo {year}
  {2016})}\BibitemShut {NoStop}%
\bibitem [{\citenamefont {Aguado}(2017)}]{Aguado2017}%
  \BibitemOpen
  \bibfield  {author} {\bibinfo {author} {\bibfnamefont {R.}~\bibnamefont
  {Aguado}},\ }\href {https://doi.org/10.1393/ncr/i2017-10141-9} {\bibfield
  {journal} {\bibinfo  {journal} {Riv. Nuovo Cimento}\ }\textbf {\bibinfo
  {volume} {40}},\ \bibinfo {pages} {523–593} (\bibinfo {year}
  {2017})}\BibitemShut {NoStop}%
\bibitem [{\citenamefont {Stanescu}\ and\ \citenamefont
  {Tewari}(2013)}]{Stanescu2013}%
  \BibitemOpen
  \bibfield  {author} {\bibinfo {author} {\bibfnamefont {T.~D.}\ \bibnamefont
  {Stanescu}}\ and\ \bibinfo {author} {\bibfnamefont {S.}~\bibnamefont
  {Tewari}},\ }\href {https://doi.org/10.1088/0953-8984/25/23/233201}
  {\bibfield  {journal} {\bibinfo  {journal} {J. Phys.: Condens. Matter}\
  }\textbf {\bibinfo {volume} {25}},\ \bibinfo {pages} {233201} (\bibinfo
  {year} {2013})}\BibitemShut {NoStop}%
\bibitem [{\citenamefont {Lutchyn}\ \emph {et~al.}(2018)\citenamefont
  {Lutchyn}, \citenamefont {Bakkers}, \citenamefont {Kouwenhoven},
  \citenamefont {Krogstrup}, \citenamefont {Marcus},\ and\ \citenamefont
  {Oreg}}]{Lutchyn2018}%
  \BibitemOpen
  \bibfield  {author} {\bibinfo {author} {\bibfnamefont {R.~M.}\ \bibnamefont
  {Lutchyn}}, \bibinfo {author} {\bibfnamefont {E.~P. A.~M.}\ \bibnamefont
  {Bakkers}}, \bibinfo {author} {\bibfnamefont {L.~P.}\ \bibnamefont
  {Kouwenhoven}}, \bibinfo {author} {\bibfnamefont {P.}~\bibnamefont
  {Krogstrup}}, \bibinfo {author} {\bibfnamefont {C.~M.}\ \bibnamefont
  {Marcus}},\ and\ \bibinfo {author} {\bibfnamefont {Y.}~\bibnamefont {Oreg}},\
  }\href {https://doi.org/10.1038/s41578-018-0003-1} {\bibfield  {journal}
  {\bibinfo  {journal} {Nat. Rev. Mater.}\ }\textbf {\bibinfo {volume} {3}},\
  \bibinfo {pages} {52} (\bibinfo {year} {2018})}\BibitemShut {NoStop}%
\bibitem [{\citenamefont {Zhang}\ \emph {et~al.}(2019)\citenamefont {Zhang},
  \citenamefont {Liu}, \citenamefont {Wimmer},\ and\ \citenamefont
  {Kouwenhoven}}]{Zhang2019}%
  \BibitemOpen
  \bibfield  {author} {\bibinfo {author} {\bibfnamefont {H.}~\bibnamefont
  {Zhang}}, \bibinfo {author} {\bibfnamefont {D.~E.}\ \bibnamefont {Liu}},
  \bibinfo {author} {\bibfnamefont {M.}~\bibnamefont {Wimmer}},\ and\ \bibinfo
  {author} {\bibfnamefont {L.~P.}\ \bibnamefont {Kouwenhoven}},\ }\href
  {https://www.nature.com/articles/s41467-019-13133-1} {\bibfield  {journal}
  {\bibinfo  {journal} {Nat. Commun.}\ }\textbf {\bibinfo {volume} {10}}
  (\bibinfo {year} {2019})}\BibitemShut {NoStop}%
\bibitem [{\citenamefont {Mourik}\ \emph {et~al.}(2012)\citenamefont {Mourik},
  \citenamefont {Zuo}, \citenamefont {Frolov}, \citenamefont {Plissard},
  \citenamefont {Bakkers},\ and\ \citenamefont {Kouwenhoven}}]{Mourik2012}%
  \BibitemOpen
  \bibfield  {author} {\bibinfo {author} {\bibfnamefont {V.}~\bibnamefont
  {Mourik}}, \bibinfo {author} {\bibfnamefont {K.}~\bibnamefont {Zuo}},
  \bibinfo {author} {\bibfnamefont {S.~M.}\ \bibnamefont {Frolov}}, \bibinfo
  {author} {\bibfnamefont {S.~R.}\ \bibnamefont {Plissard}}, \bibinfo {author}
  {\bibfnamefont {E.~P. A.~M.}\ \bibnamefont {Bakkers}},\ and\ \bibinfo
  {author} {\bibfnamefont {L.~P.}\ \bibnamefont {Kouwenhoven}},\ }\href
  {https://doi.org/10.1126/science.1222360} {\bibfield  {journal} {\bibinfo
  {journal} {Science}\ }\textbf {\bibinfo {volume} {336}},\ \bibinfo {pages}
  {1003} (\bibinfo {year} {2012})}\BibitemShut {NoStop}%
\bibitem [{\citenamefont {Kwon}\ \emph {et~al.}(2003)\citenamefont {Kwon},
  \citenamefont {Sengupta},\ and\ \citenamefont {Yakovenko}}]{Kwon2003}%
  \BibitemOpen
  \bibfield  {author} {\bibinfo {author} {\bibfnamefont {H.-J.}\ \bibnamefont
  {Kwon}}, \bibinfo {author} {\bibfnamefont {K.}~\bibnamefont {Sengupta}},\
  and\ \bibinfo {author} {\bibfnamefont {V.~M.}\ \bibnamefont {Yakovenko}},\
  }\href {https://doi.org/10.1140/epjb/e2004-00066-4} {\bibfield  {journal}
  {\bibinfo  {journal} {Eur. Phys. J. B}\ }\textbf {\bibinfo {volume} {37}},\
  \bibinfo {pages} {349} (\bibinfo {year} {2003})}\BibitemShut {NoStop}%
\bibitem [{\citenamefont {Kwon}\ \emph {et~al.}(2004)\citenamefont {Kwon},
  \citenamefont {Yakovenko},\ and\ \citenamefont {Sengupta}}]{Kwon2004}%
  \BibitemOpen
  \bibfield  {author} {\bibinfo {author} {\bibfnamefont {H.-J.}\ \bibnamefont
  {Kwon}}, \bibinfo {author} {\bibfnamefont {V.~M.}\ \bibnamefont
  {Yakovenko}},\ and\ \bibinfo {author} {\bibfnamefont {K.}~\bibnamefont
  {Sengupta}},\ }\href {https://doi.org/10.1063/1.1789931} {\bibfield
  {journal} {\bibinfo  {journal} {Low Temp. Phys.}\ }\textbf {\bibinfo {volume}
  {30}},\ \bibinfo {pages} {613} (\bibinfo {year} {2004})}\BibitemShut
  {NoStop}%
\bibitem [{\citenamefont {Fu}\ and\ \citenamefont {Kane}(2009)}]{Fu2009}%
  \BibitemOpen
  \bibfield  {author} {\bibinfo {author} {\bibfnamefont {L.}~\bibnamefont
  {Fu}}\ and\ \bibinfo {author} {\bibfnamefont {C.~L.}\ \bibnamefont {Kane}},\
  }\href {https://doi.org/10.1103/PhysRevB.79.161408} {\bibfield  {journal}
  {\bibinfo  {journal} {Phys. Rev. B}\ }\textbf {\bibinfo {volume} {79}},\
  \bibinfo {pages} {161408} (\bibinfo {year} {2009})}\BibitemShut {NoStop}%
\bibitem [{\citenamefont {Jiang}\ \emph {et~al.}(2011)\citenamefont {Jiang},
  \citenamefont {Pekker}, \citenamefont {Alicea}, \citenamefont {Refael},
  \citenamefont {Oreg},\ and\ \citenamefont {von Oppen}}]{Jiang2011}%
  \BibitemOpen
  \bibfield  {author} {\bibinfo {author} {\bibfnamefont {L.}~\bibnamefont
  {Jiang}}, \bibinfo {author} {\bibfnamefont {D.}~\bibnamefont {Pekker}},
  \bibinfo {author} {\bibfnamefont {J.}~\bibnamefont {Alicea}}, \bibinfo
  {author} {\bibfnamefont {G.}~\bibnamefont {Refael}}, \bibinfo {author}
  {\bibfnamefont {Y.}~\bibnamefont {Oreg}},\ and\ \bibinfo {author}
  {\bibfnamefont {F.}~\bibnamefont {von Oppen}},\ }\href
  {https://doi.org/10.1103/PhysRevLett.107.236401} {\bibfield  {journal}
  {\bibinfo  {journal} {Phys. Rev. Lett.}\ }\textbf {\bibinfo {volume} {107}},\
  \bibinfo {pages} {236401} (\bibinfo {year} {2011})}\BibitemShut {NoStop}%
\bibitem [{\citenamefont {Bolech}\ and\ \citenamefont
  {Demler}(2007)}]{Bolech2007}%
  \BibitemOpen
  \bibfield  {author} {\bibinfo {author} {\bibfnamefont {C.~J.}\ \bibnamefont
  {Bolech}}\ and\ \bibinfo {author} {\bibfnamefont {E.}~\bibnamefont
  {Demler}},\ }\href {https://doi.org/10.1103/PhysRevLett.98.237002} {\bibfield
   {journal} {\bibinfo  {journal} {Phys. Rev. Lett.}\ }\textbf {\bibinfo
  {volume} {98}},\ \bibinfo {pages} {237002} (\bibinfo {year}
  {2007})}\BibitemShut {NoStop}%
\bibitem [{\citenamefont {Nilsson}\ \emph {et~al.}(2008)\citenamefont
  {Nilsson}, \citenamefont {Akhmerov},\ and\ \citenamefont
  {Beenakker}}]{Nilsson2008}%
  \BibitemOpen
  \bibfield  {author} {\bibinfo {author} {\bibfnamefont {J.}~\bibnamefont
  {Nilsson}}, \bibinfo {author} {\bibfnamefont {A.~R.}\ \bibnamefont
  {Akhmerov}},\ and\ \bibinfo {author} {\bibfnamefont {C.~W.~J.}\ \bibnamefont
  {Beenakker}},\ }\href {https://doi.org/10.1103/PhysRevLett.101.120403}
  {\bibfield  {journal} {\bibinfo  {journal} {Phys. Rev. Lett.}\ }\textbf
  {\bibinfo {volume} {101}},\ \bibinfo {pages} {120403} (\bibinfo {year}
  {2008})}\BibitemShut {NoStop}%
\bibitem [{\citenamefont {Law}\ \emph {et~al.}(2009)\citenamefont {Law},
  \citenamefont {Lee},\ and\ \citenamefont {Ng}}]{Law2009}%
  \BibitemOpen
  \bibfield  {author} {\bibinfo {author} {\bibfnamefont {K.~T.}\ \bibnamefont
  {Law}}, \bibinfo {author} {\bibfnamefont {P.~A.}\ \bibnamefont {Lee}},\ and\
  \bibinfo {author} {\bibfnamefont {T.~K.}\ \bibnamefont {Ng}},\ }\href
  {https://doi.org/10.1103/PhysRevLett.103.237001} {\bibfield  {journal}
  {\bibinfo  {journal} {Phys. Rev. Lett.}\ }\textbf {\bibinfo {volume} {103}},\
  \bibinfo {pages} {237001} (\bibinfo {year} {2009})}\BibitemShut {NoStop}%
\bibitem [{\citenamefont {Golub}\ and\ \citenamefont
  {Horovitz}(2011)}]{Golub2011}%
  \BibitemOpen
  \bibfield  {author} {\bibinfo {author} {\bibfnamefont {A.}~\bibnamefont
  {Golub}}\ and\ \bibinfo {author} {\bibfnamefont {B.}~\bibnamefont
  {Horovitz}},\ }\href {https://doi.org/10.1103/PhysRevB.83.153415} {\bibfield
  {journal} {\bibinfo  {journal} {Phys. Rev. B}\ }\textbf {\bibinfo {volume}
  {83}},\ \bibinfo {pages} {153415} (\bibinfo {year} {2011})}\BibitemShut
  {NoStop}%
\bibitem [{\citenamefont {Kells}\ \emph {et~al.}(2012)\citenamefont {Kells},
  \citenamefont {Meidan},\ and\ \citenamefont {Brouwer}}]{Kells2012}%
  \BibitemOpen
  \bibfield  {author} {\bibinfo {author} {\bibfnamefont {G.}~\bibnamefont
  {Kells}}, \bibinfo {author} {\bibfnamefont {D.}~\bibnamefont {Meidan}},\ and\
  \bibinfo {author} {\bibfnamefont {P.~W.}\ \bibnamefont {Brouwer}},\ }\href
  {https://doi.org/10.1103/PhysRevB.86.100503} {\bibfield  {journal} {\bibinfo
  {journal} {Phys. Rev. B}\ }\textbf {\bibinfo {volume} {86}},\ \bibinfo
  {pages} {100503} (\bibinfo {year} {2012})}\BibitemShut {NoStop}%
\bibitem [{\citenamefont {Prada}\ \emph {et~al.}(2012)\citenamefont {Prada},
  \citenamefont {San-Jose},\ and\ \citenamefont {Aguado}}]{Prada2012}%
  \BibitemOpen
  \bibfield  {author} {\bibinfo {author} {\bibfnamefont {E.}~\bibnamefont
  {Prada}}, \bibinfo {author} {\bibfnamefont {P.}~\bibnamefont {San-Jose}},\
  and\ \bibinfo {author} {\bibfnamefont {R.}~\bibnamefont {Aguado}},\ }\href
  {https://doi.org/10.1103/PhysRevB.86.180503} {\bibfield  {journal} {\bibinfo
  {journal} {Phys. Rev. B}\ }\textbf {\bibinfo {volume} {86}},\ \bibinfo
  {pages} {180503} (\bibinfo {year} {2012})}\BibitemShut {NoStop}%
\bibitem [{\citenamefont {Cayao}\ \emph {et~al.}(2015)\citenamefont {Cayao},
  \citenamefont {Prada}, \citenamefont {San-Jose},\ and\ \citenamefont
  {Aguado}}]{Cayao2015}%
  \BibitemOpen
  \bibfield  {author} {\bibinfo {author} {\bibfnamefont {J.}~\bibnamefont
  {Cayao}}, \bibinfo {author} {\bibfnamefont {E.}~\bibnamefont {Prada}},
  \bibinfo {author} {\bibfnamefont {P.}~\bibnamefont {San-Jose}},\ and\
  \bibinfo {author} {\bibfnamefont {R.}~\bibnamefont {Aguado}},\ }\href
  {https://doi.org/10.1103/PhysRevB.91.024514} {\bibfield  {journal} {\bibinfo
  {journal} {Phys. Rev. B}\ }\textbf {\bibinfo {volume} {91}},\ \bibinfo
  {pages} {024514} (\bibinfo {year} {2015})}\BibitemShut {NoStop}%
\bibitem [{\citenamefont {Liu}\ \emph {et~al.}(2017)\citenamefont {Liu},
  \citenamefont {Sau}, \citenamefont {Stanescu},\ and\ \citenamefont
  {Das~Sarma}}]{Liu2017}%
  \BibitemOpen
  \bibfield  {author} {\bibinfo {author} {\bibfnamefont {C.-X.}\ \bibnamefont
  {Liu}}, \bibinfo {author} {\bibfnamefont {J.~D.}\ \bibnamefont {Sau}},
  \bibinfo {author} {\bibfnamefont {T.~D.}\ \bibnamefont {Stanescu}},\ and\
  \bibinfo {author} {\bibfnamefont {S.}~\bibnamefont {Das~Sarma}},\ }\href
  {https://doi.org/10.1103/PhysRevB.96.075161} {\bibfield  {journal} {\bibinfo
  {journal} {Phys. Rev. B}\ }\textbf {\bibinfo {volume} {96}},\ \bibinfo
  {pages} {075161} (\bibinfo {year} {2017})}\BibitemShut {NoStop}%
\bibitem [{\citenamefont {Moore}\ \emph
  {et~al.}(2018{\natexlab{a}})\citenamefont {Moore}, \citenamefont {Stanescu},\
  and\ \citenamefont {Tewari}}]{Moore2018}%
  \BibitemOpen
  \bibfield  {author} {\bibinfo {author} {\bibfnamefont {C.}~\bibnamefont
  {Moore}}, \bibinfo {author} {\bibfnamefont {T.~D.}\ \bibnamefont
  {Stanescu}},\ and\ \bibinfo {author} {\bibfnamefont {S.}~\bibnamefont
  {Tewari}},\ }\href {https://doi.org/10.1103/PhysRevB.97.165302} {\bibfield
  {journal} {\bibinfo  {journal} {Phys. Rev. B}\ }\textbf {\bibinfo {volume}
  {97}},\ \bibinfo {pages} {165302} (\bibinfo {year}
  {2018}{\natexlab{a}})}\BibitemShut {NoStop}%
\bibitem [{\citenamefont {Moore}\ \emph
  {et~al.}(2018{\natexlab{b}})\citenamefont {Moore}, \citenamefont {Zeng},
  \citenamefont {Stanescu},\ and\ \citenamefont {Tewari}}]{Moore2018a}%
  \BibitemOpen
  \bibfield  {author} {\bibinfo {author} {\bibfnamefont {C.}~\bibnamefont
  {Moore}}, \bibinfo {author} {\bibfnamefont {C.}~\bibnamefont {Zeng}},
  \bibinfo {author} {\bibfnamefont {T.~D.}\ \bibnamefont {Stanescu}},\ and\
  \bibinfo {author} {\bibfnamefont {S.}~\bibnamefont {Tewari}},\ }\href
  {https://doi.org/10.1103/PhysRevB.98.155314} {\bibfield  {journal} {\bibinfo
  {journal} {Phys. Rev. B}\ }\textbf {\bibinfo {volume} {98}},\ \bibinfo
  {pages} {155314} (\bibinfo {year} {2018}{\natexlab{b}})}\BibitemShut
  {NoStop}%
\bibitem [{\citenamefont {Awoga}\ \emph {et~al.}(2019)\citenamefont {Awoga},
  \citenamefont {Cayao},\ and\ \citenamefont {Black-Schaffer}}]{Awoga2019}%
  \BibitemOpen
  \bibfield  {author} {\bibinfo {author} {\bibfnamefont {O.~A.}\ \bibnamefont
  {Awoga}}, \bibinfo {author} {\bibfnamefont {J.}~\bibnamefont {Cayao}},\ and\
  \bibinfo {author} {\bibfnamefont {A.~M.}\ \bibnamefont {Black-Schaffer}},\
  }\href {https://doi.org/10.1103/PhysRevLett.123.117001} {\bibfield  {journal}
  {\bibinfo  {journal} {Phys. Rev. Lett.}\ }\textbf {\bibinfo {volume} {123}},\
  \bibinfo {pages} {117001} (\bibinfo {year} {2019})}\BibitemShut {NoStop}%
\bibitem [{\citenamefont {Vuik}\ \emph {et~al.}(2019)\citenamefont {Vuik},
  \citenamefont {Nijholt}, \citenamefont {Akhmerov},\ and\ \citenamefont
  {Wimmer}}]{Vuik2019}%
  \BibitemOpen
  \bibfield  {author} {\bibinfo {author} {\bibfnamefont {A.}~\bibnamefont
  {Vuik}}, \bibinfo {author} {\bibfnamefont {B.}~\bibnamefont {Nijholt}},
  \bibinfo {author} {\bibfnamefont {A.~R.}\ \bibnamefont {Akhmerov}},\ and\
  \bibinfo {author} {\bibfnamefont {M.}~\bibnamefont {Wimmer}},\ }\href
  {https://doi.org/10.21468/SciPostPhys.7.5.061} {\bibfield  {journal}
  {\bibinfo  {journal} {SciPost Phys.}\ }\textbf {\bibinfo {volume} {7}},\
  \bibinfo {pages} {61} (\bibinfo {year} {2019})}\BibitemShut {NoStop}%
\bibitem [{\citenamefont {Ojanen}(2012)}]{Ojanen2012}%
  \BibitemOpen
  \bibfield  {author} {\bibinfo {author} {\bibfnamefont {T.}~\bibnamefont
  {Ojanen}},\ }\href {https://doi.org/10.1103/PhysRevLett.109.226804}
  {\bibfield  {journal} {\bibinfo  {journal} {Phys. Rev. Lett.}\ }\textbf
  {\bibinfo {volume} {109}},\ \bibinfo {pages} {226804} (\bibinfo {year}
  {2012})}\BibitemShut {NoStop}%
\bibitem [{\citenamefont {Ojanen}\ and\ \citenamefont
  {Kitagawa}(2013)}]{Ojanen2013}%
  \BibitemOpen
  \bibfield  {author} {\bibinfo {author} {\bibfnamefont {T.}~\bibnamefont
  {Ojanen}}\ and\ \bibinfo {author} {\bibfnamefont {T.}~\bibnamefont
  {Kitagawa}},\ }\href {https://doi.org/10.1103/PhysRevB.87.014512} {\bibfield
  {journal} {\bibinfo  {journal} {Phys. Rev. B}\ }\textbf {\bibinfo {volume}
  {87}},\ \bibinfo {pages} {014512} (\bibinfo {year} {2013})}\BibitemShut
  {NoStop}%
\bibitem [{\citenamefont {Oliveira}\ and\ \citenamefont
  {Sacramento}(2014)}]{Oliveira2014}%
  \BibitemOpen
  \bibfield  {author} {\bibinfo {author} {\bibfnamefont {T.~P.}\ \bibnamefont
  {Oliveira}}\ and\ \bibinfo {author} {\bibfnamefont {P.~D.}\ \bibnamefont
  {Sacramento}},\ }\href {https://doi.org/10.1103/PhysRevB.89.094512}
  {\bibfield  {journal} {\bibinfo  {journal} {Phys. Rev. B}\ }\textbf {\bibinfo
  {volume} {89}},\ \bibinfo {pages} {094512} (\bibinfo {year}
  {2014})}\BibitemShut {NoStop}%
\bibitem [{\citenamefont {Setiawan}\ \emph {et~al.}(2015)\citenamefont
  {Setiawan}, \citenamefont {Sengupta}, \citenamefont {Spielman},\ and\
  \citenamefont {Sau}}]{Setiawan2015}%
  \BibitemOpen
  \bibfield  {author} {\bibinfo {author} {\bibfnamefont {F.}~\bibnamefont
  {Setiawan}}, \bibinfo {author} {\bibfnamefont {K.}~\bibnamefont {Sengupta}},
  \bibinfo {author} {\bibfnamefont {I.~B.}\ \bibnamefont {Spielman}},\ and\
  \bibinfo {author} {\bibfnamefont {J.~D.}\ \bibnamefont {Sau}},\ }\href
  {https://doi.org/10.1103/PhysRevLett.115.190401} {\bibfield  {journal}
  {\bibinfo  {journal} {Phys. Rev. Lett.}\ }\textbf {\bibinfo {volume} {115}},\
  \bibinfo {pages} {190401} (\bibinfo {year} {2015})}\BibitemShut {NoStop}%
\bibitem [{\citenamefont {Cayao}\ \emph {et~al.}(2017)\citenamefont {Cayao},
  \citenamefont {San-Jose}, \citenamefont {Black-Schaffer}, \citenamefont
  {Aguado},\ and\ \citenamefont {Prada}}]{Cayao2017}%
  \BibitemOpen
  \bibfield  {author} {\bibinfo {author} {\bibfnamefont {J.}~\bibnamefont
  {Cayao}}, \bibinfo {author} {\bibfnamefont {P.}~\bibnamefont {San-Jose}},
  \bibinfo {author} {\bibfnamefont {A.~M.}\ \bibnamefont {Black-Schaffer}},
  \bibinfo {author} {\bibfnamefont {R.}~\bibnamefont {Aguado}},\ and\ \bibinfo
  {author} {\bibfnamefont {E.}~\bibnamefont {Prada}},\ }\href
  {https://doi.org/10.1103/PhysRevB.96.205425} {\bibfield  {journal} {\bibinfo
  {journal} {Phys. Rev. B}\ }\textbf {\bibinfo {volume} {96}},\ \bibinfo
  {pages} {205425} (\bibinfo {year} {2017})}\BibitemShut {NoStop}%
\bibitem [{\citenamefont {Rosdahl}\ \emph {et~al.}(2018)\citenamefont
  {Rosdahl}, \citenamefont {Vuik}, \citenamefont {Kjaergaard},\ and\
  \citenamefont {Akhmerov}}]{Rosdahl2018}%
  \BibitemOpen
  \bibfield  {author} {\bibinfo {author} {\bibfnamefont {T.~O.}\ \bibnamefont
  {Rosdahl}}, \bibinfo {author} {\bibfnamefont {A.}~\bibnamefont {Vuik}},
  \bibinfo {author} {\bibfnamefont {M.}~\bibnamefont {Kjaergaard}},\ and\
  \bibinfo {author} {\bibfnamefont {A.~R.}\ \bibnamefont {Akhmerov}},\ }\href
  {https://doi.org/10.1103/PhysRevB.97.045421} {\bibfield  {journal} {\bibinfo
  {journal} {Phys. Rev. B}\ }\textbf {\bibinfo {volume} {97}},\ \bibinfo
  {pages} {045421} (\bibinfo {year} {2018})}\BibitemShut {NoStop}%
\bibitem [{\citenamefont {Szumniak}\ \emph {et~al.}(2017)\citenamefont
  {Szumniak}, \citenamefont {Chevallier}, \citenamefont {Loss},\ and\
  \citenamefont {Klinovaja}}]{Szumniak2017}%
  \BibitemOpen
  \bibfield  {author} {\bibinfo {author} {\bibfnamefont {P.}~\bibnamefont
  {Szumniak}}, \bibinfo {author} {\bibfnamefont {D.}~\bibnamefont
  {Chevallier}}, \bibinfo {author} {\bibfnamefont {D.}~\bibnamefont {Loss}},\
  and\ \bibinfo {author} {\bibfnamefont {J.}~\bibnamefont {Klinovaja}},\ }\href
  {https://doi.org/10.1103/PhysRevB.96.041401} {\bibfield  {journal} {\bibinfo
  {journal} {Phys. Rev. B}\ }\textbf {\bibinfo {volume} {96}},\ \bibinfo
  {pages} {041401} (\bibinfo {year} {2017})}\BibitemShut {NoStop}%
\bibitem [{\citenamefont {Serina}\ \emph {et~al.}(2018)\citenamefont {Serina},
  \citenamefont {Loss},\ and\ \citenamefont {Klinovaja}}]{Serina2018}%
  \BibitemOpen
  \bibfield  {author} {\bibinfo {author} {\bibfnamefont {M.}~\bibnamefont
  {Serina}}, \bibinfo {author} {\bibfnamefont {D.}~\bibnamefont {Loss}},\ and\
  \bibinfo {author} {\bibfnamefont {J.}~\bibnamefont {Klinovaja}},\ }\href
  {https://doi.org/10.1103/PhysRevB.98.035419} {\bibfield  {journal} {\bibinfo
  {journal} {Phys. Rev. B}\ }\textbf {\bibinfo {volume} {98}},\ \bibinfo
  {pages} {035419} (\bibinfo {year} {2018})}\BibitemShut {NoStop}%
\bibitem [{\citenamefont {Chen}\ \emph {et~al.}(2019)\citenamefont {Chen},
  \citenamefont {Wu},\ and\ \citenamefont {Liu}}]{Chen2019}%
  \BibitemOpen
  \bibfield  {author} {\bibinfo {author} {\bibfnamefont {L.}~\bibnamefont
  {Chen}}, \bibinfo {author} {\bibfnamefont {Y.-H.}\ \bibnamefont {Wu}},\ and\
  \bibinfo {author} {\bibfnamefont {X.}~\bibnamefont {Liu}},\ }\href
  {https://doi.org/10.1103/PhysRevB.99.165307} {\bibfield  {journal} {\bibinfo
  {journal} {Phys. Rev. B}\ }\textbf {\bibinfo {volume} {99}},\ \bibinfo
  {pages} {165307} (\bibinfo {year} {2019})}\BibitemShut {NoStop}%
\bibitem [{\citenamefont {Chevallier}\ \emph {et~al.}(2018)\citenamefont
  {Chevallier}, \citenamefont {Szumniak}, \citenamefont {Hoffman},
  \citenamefont {Loss},\ and\ \citenamefont {Klinovaja}}]{Chevallier2018}%
  \BibitemOpen
  \bibfield  {author} {\bibinfo {author} {\bibfnamefont {D.}~\bibnamefont
  {Chevallier}}, \bibinfo {author} {\bibfnamefont {P.}~\bibnamefont
  {Szumniak}}, \bibinfo {author} {\bibfnamefont {S.}~\bibnamefont {Hoffman}},
  \bibinfo {author} {\bibfnamefont {D.}~\bibnamefont {Loss}},\ and\ \bibinfo
  {author} {\bibfnamefont {J.}~\bibnamefont {Klinovaja}},\ }\href
  {https://doi.org/10.1103/PhysRevB.97.045404} {\bibfield  {journal} {\bibinfo
  {journal} {Phys. Rev. B}\ }\textbf {\bibinfo {volume} {97}},\ \bibinfo
  {pages} {045404} (\bibinfo {year} {2018})}\BibitemShut {NoStop}%
\bibitem [{\citenamefont {J\"{u}nger}\ \emph {et~al.}(2019)\citenamefont
  {J\"{u}nger}, \citenamefont {Baumgartner}, \citenamefont {Delagrange},
  \citenamefont {Chevallier}, \citenamefont {Lehmann}, \citenamefont {Nilsson},
  \citenamefont {Dick}, \citenamefont {Thelander},\ and\ \citenamefont
  {Sch\"{o}nenberger}}]{Juenger2019}%
  \BibitemOpen
  \bibfield  {author} {\bibinfo {author} {\bibfnamefont {C.}~\bibnamefont
  {J\"{u}nger}}, \bibinfo {author} {\bibfnamefont {A.}~\bibnamefont
  {Baumgartner}}, \bibinfo {author} {\bibfnamefont {R.}~\bibnamefont
  {Delagrange}}, \bibinfo {author} {\bibfnamefont {D.}~\bibnamefont
  {Chevallier}}, \bibinfo {author} {\bibfnamefont {S.}~\bibnamefont {Lehmann}},
  \bibinfo {author} {\bibfnamefont {M.}~\bibnamefont {Nilsson}}, \bibinfo
  {author} {\bibfnamefont {K.~A.}\ \bibnamefont {Dick}}, \bibinfo {author}
  {\bibfnamefont {C.}~\bibnamefont {Thelander}},\ and\ \bibinfo {author}
  {\bibfnamefont {C.}~\bibnamefont {Sch\"{o}nenberger}},\ }\href
  {https://doi.org/10.1038/s42005-019-0162-4} {\bibfield  {journal} {\bibinfo
  {journal} {Commun. Phys.}\ }\textbf {\bibinfo {volume} {2}} (\bibinfo {year}
  {2019})}\BibitemShut {NoStop}%
\bibitem [{\citenamefont {Slichter}(1990)}]{Slichter1990}%
  \BibitemOpen
  \bibfield  {author} {\bibinfo {author} {\bibfnamefont {C.~P.}\ \bibnamefont
  {Slichter}},\ }\href {https://doi.org/10.1007/978-3-662-09441-9} {\emph
  {\bibinfo {title} {Principles of Magnetic Resonance}}}\ (\bibinfo
  {publisher} {Springer Berlin Heidelberg},\ \bibinfo {year}
  {1990})\BibitemShut {NoStop}%
\bibitem [{\citenamefont {Nitta}\ \emph {et~al.}(1997)\citenamefont {Nitta},
  \citenamefont {Akazaki}, \citenamefont {Takayanagi},\ and\ \citenamefont
  {Enoki}}]{Nitta1997}%
  \BibitemOpen
  \bibfield  {author} {\bibinfo {author} {\bibfnamefont {J.}~\bibnamefont
  {Nitta}}, \bibinfo {author} {\bibfnamefont {T.}~\bibnamefont {Akazaki}},
  \bibinfo {author} {\bibfnamefont {H.}~\bibnamefont {Takayanagi}},\ and\
  \bibinfo {author} {\bibfnamefont {T.}~\bibnamefont {Enoki}},\ }\href
  {https://doi.org/10.1103/PhysRevLett.78.1335} {\bibfield  {journal} {\bibinfo
   {journal} {Phys. Rev. Lett.}\ }\textbf {\bibinfo {volume} {78}},\ \bibinfo
  {pages} {1335} (\bibinfo {year} {1997})}\BibitemShut {NoStop}%
\bibitem [{\citenamefont {Salis}\ \emph {et~al.}(2001)\citenamefont {Salis},
  \citenamefont {Kato}, \citenamefont {Ensslin}, \citenamefont {Driscoll},
  \citenamefont {Gossard},\ and\ \citenamefont {Awschalom}}]{Salis2001}%
  \BibitemOpen
  \bibfield  {author} {\bibinfo {author} {\bibfnamefont {G.}~\bibnamefont
  {Salis}}, \bibinfo {author} {\bibfnamefont {Y.}~\bibnamefont {Kato}},
  \bibinfo {author} {\bibfnamefont {K.}~\bibnamefont {Ensslin}}, \bibinfo
  {author} {\bibfnamefont {D.~C.}\ \bibnamefont {Driscoll}}, \bibinfo {author}
  {\bibfnamefont {A.~C.}\ \bibnamefont {Gossard}},\ and\ \bibinfo {author}
  {\bibfnamefont {D.~D.}\ \bibnamefont {Awschalom}},\ }\href
  {https://doi.org/10.1038/414619a} {\bibfield  {journal} {\bibinfo  {journal}
  {Nature}\ }\textbf {\bibinfo {volume} {414}},\ \bibinfo {pages} {619}
  (\bibinfo {year} {2001})}\BibitemShut {NoStop}%
\bibitem [{\citenamefont {Kato}\ \emph {et~al.}(2003)\citenamefont {Kato},
  \citenamefont {Myers}, \citenamefont {Driscoll}, \citenamefont {Gossard},
  \citenamefont {Levy},\ and\ \citenamefont {Awschalom}}]{Kato2003}%
  \BibitemOpen
  \bibfield  {author} {\bibinfo {author} {\bibfnamefont {Y.}~\bibnamefont
  {Kato}}, \bibinfo {author} {\bibfnamefont {R.~C.}\ \bibnamefont {Myers}},
  \bibinfo {author} {\bibfnamefont {D.~C.}\ \bibnamefont {Driscoll}}, \bibinfo
  {author} {\bibfnamefont {A.~C.}\ \bibnamefont {Gossard}}, \bibinfo {author}
  {\bibfnamefont {J.}~\bibnamefont {Levy}},\ and\ \bibinfo {author}
  {\bibfnamefont {D.~D.}\ \bibnamefont {Awschalom}},\ }\href
  {https://doi.org/10.1126/science.1080880} {\bibfield  {journal} {\bibinfo
  {journal} {Science}\ }\textbf {\bibinfo {volume} {299}},\ \bibinfo {pages}
  {1201} (\bibinfo {year} {2003})}\BibitemShut {NoStop}%
\bibitem [{\citenamefont {Kato}\ \emph {et~al.}(2004)\citenamefont {Kato},
  \citenamefont {Myers}, \citenamefont {Gossard},\ and\ \citenamefont
  {Awschalom}}]{Kato2004}%
  \BibitemOpen
  \bibfield  {author} {\bibinfo {author} {\bibfnamefont {Y.}~\bibnamefont
  {Kato}}, \bibinfo {author} {\bibfnamefont {R.~C.}\ \bibnamefont {Myers}},
  \bibinfo {author} {\bibfnamefont {A.~C.}\ \bibnamefont {Gossard}},\ and\
  \bibinfo {author} {\bibfnamefont {D.~D.}\ \bibnamefont {Awschalom}},\ }\href
  {https://doi.org/10.1038/nature02202} {\bibfield  {journal} {\bibinfo
  {journal} {Nature}\ }\textbf {\bibinfo {volume} {427}},\ \bibinfo {pages}
  {50} (\bibinfo {year} {2004})}\BibitemShut {NoStop}%
\bibitem [{\citenamefont {Tang}\ \emph {et~al.}(2006)\citenamefont {Tang},
  \citenamefont {Levy},\ and\ \citenamefont {Flatt\'e}}]{Tang2006}%
  \BibitemOpen
  \bibfield  {author} {\bibinfo {author} {\bibfnamefont {J.-M.}\ \bibnamefont
  {Tang}}, \bibinfo {author} {\bibfnamefont {J.}~\bibnamefont {Levy}},\ and\
  \bibinfo {author} {\bibfnamefont {M.~E.}\ \bibnamefont {Flatt\'e}},\ }\href
  {https://doi.org/10.1103/PhysRevLett.97.106803} {\bibfield  {journal}
  {\bibinfo  {journal} {Phys. Rev. Lett.}\ }\textbf {\bibinfo {volume} {97}},\
  \bibinfo {pages} {106803} (\bibinfo {year} {2006})}\BibitemShut {NoStop}%
\bibitem [{\citenamefont {Nowack}\ \emph {et~al.}(2007)\citenamefont {Nowack},
  \citenamefont {Koppens}, \citenamefont {Nazarov},\ and\ \citenamefont
  {Vandersypen}}]{Nowack2007}%
  \BibitemOpen
  \bibfield  {author} {\bibinfo {author} {\bibfnamefont {K.~C.}\ \bibnamefont
  {Nowack}}, \bibinfo {author} {\bibfnamefont {F.~H.~L.}\ \bibnamefont
  {Koppens}}, \bibinfo {author} {\bibfnamefont {Y.~V.}\ \bibnamefont
  {Nazarov}},\ and\ \bibinfo {author} {\bibfnamefont {L.~M.~K.}\ \bibnamefont
  {Vandersypen}},\ }\href {https://doi.org/10.1126/science.1148092} {\bibfield
  {journal} {\bibinfo  {journal} {Science}\ }\textbf {\bibinfo {volume}
  {318}},\ \bibinfo {pages} {1430} (\bibinfo {year} {2007})}\BibitemShut
  {NoStop}%
\bibitem [{\citenamefont {\ifmmode \check{C}\else
  \v{C}\fi{}ade\ifmmode~\check{z}\else \v{z}\fi{}}\ \emph
  {et~al.}(2014)\citenamefont {\ifmmode \check{C}\else
  \v{C}\fi{}ade\ifmmode~\check{z}\else \v{z}\fi{}}, \citenamefont {Jefferson},\
  and\ \citenamefont {Ram\ifmmode~\check{s}\else \v{s}\fi{}ak}}]{Jeff2014}%
  \BibitemOpen
  \bibfield  {author} {\bibinfo {author} {\bibfnamefont {T.}~\bibnamefont
  {\ifmmode \check{C}\else \v{C}\fi{}ade\ifmmode~\check{z}\else \v{z}\fi{}}},
  \bibinfo {author} {\bibfnamefont {J.~H.}\ \bibnamefont {Jefferson}},\ and\
  \bibinfo {author} {\bibfnamefont {A.}~\bibnamefont
  {Ram\ifmmode~\check{s}\else \v{s}\fi{}ak}},\ }\href
  {https://doi.org/10.1103/PhysRevLett.112.150402} {\bibfield  {journal}
  {\bibinfo  {journal} {Phys. Rev. Lett.}\ }\textbf {\bibinfo {volume} {112}},\
  \bibinfo {pages} {150402} (\bibinfo {year} {2014})}\BibitemShut {NoStop}%
\bibitem [{\citenamefont {Paw{\l}owski}\ \emph {et~al.}(2014)\citenamefont
  {Paw{\l}owski}, \citenamefont {Szumniak}, \citenamefont {Skubis},\ and\
  \citenamefont {Bednarek}}]{Pawlowski2014}%
  \BibitemOpen
  \bibfield  {author} {\bibinfo {author} {\bibfnamefont {J.}~\bibnamefont
  {Paw{\l}owski}}, \bibinfo {author} {\bibfnamefont {P.}~\bibnamefont
  {Szumniak}}, \bibinfo {author} {\bibfnamefont {A.}~\bibnamefont {Skubis}},\
  and\ \bibinfo {author} {\bibfnamefont {S.}~\bibnamefont {Bednarek}},\ }\href
  {https://doi.org/10.1088/0953-8984/26/34/345302} {\bibfield  {journal}
  {\bibinfo  {journal} {J. Phys.: Condens. Matter}\ }\textbf {\bibinfo {volume}
  {26}},\ \bibinfo {pages} {345302} (\bibinfo {year} {2014})}\BibitemShut
  {NoStop}%
\bibitem [{\citenamefont {Paw\l{}owski}\ \emph {et~al.}(2016)\citenamefont
  {Paw\l{}owski}, \citenamefont {Szumniak},\ and\ \citenamefont
  {Bednarek}}]{Pawlowski2016}%
  \BibitemOpen
  \bibfield  {author} {\bibinfo {author} {\bibfnamefont {J.}~\bibnamefont
  {Paw\l{}owski}}, \bibinfo {author} {\bibfnamefont {P.}~\bibnamefont
  {Szumniak}},\ and\ \bibinfo {author} {\bibfnamefont {S.}~\bibnamefont
  {Bednarek}},\ }\href {https://doi.org/10.1103/PhysRevB.94.155407} {\bibfield
  {journal} {\bibinfo  {journal} {Phys. Rev. B}\ }\textbf {\bibinfo {volume}
  {94}},\ \bibinfo {pages} {155407} (\bibinfo {year} {2016})}\BibitemShut
  {NoStop}%
\bibitem [{\citenamefont {Mal'shukov}\ \emph {et~al.}(2003)\citenamefont
  {Mal'shukov}, \citenamefont {Tang}, \citenamefont {Chu},\ and\ \citenamefont
  {Chao}}]{Malshukov2003}%
  \BibitemOpen
  \bibfield  {author} {\bibinfo {author} {\bibfnamefont {A.~G.}\ \bibnamefont
  {Mal'shukov}}, \bibinfo {author} {\bibfnamefont {C.~S.}\ \bibnamefont
  {Tang}}, \bibinfo {author} {\bibfnamefont {C.~S.}\ \bibnamefont {Chu}},\ and\
  \bibinfo {author} {\bibfnamefont {K.~A.}\ \bibnamefont {Chao}},\ }\href
  {https://doi.org/10.1103/PhysRevB.68.233307} {\bibfield  {journal} {\bibinfo
  {journal} {Phys. Rev. B}\ }\textbf {\bibinfo {volume} {68}},\ \bibinfo
  {pages} {233307} (\bibinfo {year} {2003})}\BibitemShut {NoStop}%
\bibitem [{\citenamefont {Tang}\ \emph {et~al.}(2005)\citenamefont {Tang},
  \citenamefont {Mal'shukov},\ and\ \citenamefont {Chao}}]{Tang2005}%
  \BibitemOpen
  \bibfield  {author} {\bibinfo {author} {\bibfnamefont {C.~S.}\ \bibnamefont
  {Tang}}, \bibinfo {author} {\bibfnamefont {A.~G.}\ \bibnamefont
  {Mal'shukov}},\ and\ \bibinfo {author} {\bibfnamefont {K.~A.}\ \bibnamefont
  {Chao}},\ }\href {https://doi.org/10.1103/PhysRevB.71.195314} {\bibfield
  {journal} {\bibinfo  {journal} {Phys. Rev. B}\ }\textbf {\bibinfo {volume}
  {71}},\ \bibinfo {pages} {195314} (\bibinfo {year} {2005})}\BibitemShut
  {NoStop}%
\bibitem [{\citenamefont {Liang}\ \emph {et~al.}(2009)\citenamefont {Liang},
  \citenamefont {Yang},\ and\ \citenamefont {Wang}}]{Liang2009}%
  \BibitemOpen
  \bibfield  {author} {\bibinfo {author} {\bibfnamefont {F.}~\bibnamefont
  {Liang}}, \bibinfo {author} {\bibfnamefont {Y.~H.}\ \bibnamefont {Yang}},\
  and\ \bibinfo {author} {\bibfnamefont {J.}~\bibnamefont {Wang}},\ }\href
  {https://doi.org/10.1140/epjb/e2009-00144-1} {\bibfield  {journal} {\bibinfo
  {journal} {Eur. Phys. J. B}\ }\textbf {\bibinfo {volume} {69}},\ \bibinfo
  {pages} {337} (\bibinfo {year} {2009})}\BibitemShut {NoStop}%
\bibitem [{\citenamefont {Ho}\ \emph {et~al.}(2014)\citenamefont {Ho},
  \citenamefont {Jalil},\ and\ \citenamefont {Tan}}]{Ho2014}%
  \BibitemOpen
  \bibfield  {author} {\bibinfo {author} {\bibfnamefont {C.~S.}\ \bibnamefont
  {Ho}}, \bibinfo {author} {\bibfnamefont {M.~B.~A.}\ \bibnamefont {Jalil}},\
  and\ \bibinfo {author} {\bibfnamefont {S.~G.}\ \bibnamefont {Tan}},\ }\href
  {https://doi.org/10.1063/1.4876226} {\bibfield  {journal} {\bibinfo
  {journal} {J. Appl. Phys.}\ }\textbf {\bibinfo {volume} {115}},\ \bibinfo
  {pages} {183705} (\bibinfo {year} {2014})}\BibitemShut {NoStop}%
\bibitem [{\citenamefont {Deng}\ \emph {et~al.}(2012)\citenamefont {Deng},
  \citenamefont {Yu}, \citenamefont {Huang}, \citenamefont {Larsson},
  \citenamefont {Caroff},\ and\ \citenamefont {Xu}}]{Deng2012}%
  \BibitemOpen
  \bibfield  {author} {\bibinfo {author} {\bibfnamefont {M.~T.}\ \bibnamefont
  {Deng}}, \bibinfo {author} {\bibfnamefont {C.~L.}\ \bibnamefont {Yu}},
  \bibinfo {author} {\bibfnamefont {G.~Y.}\ \bibnamefont {Huang}}, \bibinfo
  {author} {\bibfnamefont {M.}~\bibnamefont {Larsson}}, \bibinfo {author}
  {\bibfnamefont {P.}~\bibnamefont {Caroff}},\ and\ \bibinfo {author}
  {\bibfnamefont {H.~Q.}\ \bibnamefont {Xu}},\ }\href
  {https://doi.org/10.1021/nl303758w} {\bibfield  {journal} {\bibinfo
  {journal} {Nano Lett.}\ }\textbf {\bibinfo {volume} {12}},\ \bibinfo {pages}
  {6414} (\bibinfo {year} {2012})}\BibitemShut {NoStop}%
\bibitem [{\citenamefont {Rokhinson}\ \emph {et~al.}(2012)\citenamefont
  {Rokhinson}, \citenamefont {Liu},\ and\ \citenamefont
  {Furdyna}}]{Rokhinson2012}%
  \BibitemOpen
  \bibfield  {author} {\bibinfo {author} {\bibfnamefont {L.~P.}\ \bibnamefont
  {Rokhinson}}, \bibinfo {author} {\bibfnamefont {X.}~\bibnamefont {Liu}},\
  and\ \bibinfo {author} {\bibfnamefont {J.~K.}\ \bibnamefont {Furdyna}},\
  }\href {https://doi.org/10.1038/nphys2429} {\bibfield  {journal} {\bibinfo
  {journal} {Nat. Phys.}\ }\textbf {\bibinfo {volume} {8}},\ \bibinfo {pages}
  {795} (\bibinfo {year} {2012})}\BibitemShut {NoStop}%
\bibitem [{\citenamefont {Das}\ \emph {et~al.}(2012)\citenamefont {Das},
  \citenamefont {Ronen}, \citenamefont {Most}, \citenamefont {Oreg},
  \citenamefont {Heiblum},\ and\ \citenamefont {Shtrikman}}]{Das2012}%
  \BibitemOpen
  \bibfield  {author} {\bibinfo {author} {\bibfnamefont {A.}~\bibnamefont
  {Das}}, \bibinfo {author} {\bibfnamefont {Y.}~\bibnamefont {Ronen}}, \bibinfo
  {author} {\bibfnamefont {Y.}~\bibnamefont {Most}}, \bibinfo {author}
  {\bibfnamefont {Y.}~\bibnamefont {Oreg}}, \bibinfo {author} {\bibfnamefont
  {M.}~\bibnamefont {Heiblum}},\ and\ \bibinfo {author} {\bibfnamefont
  {H.}~\bibnamefont {Shtrikman}},\ }\href {https://doi.org/10.1038/nphys2479}
  {\bibfield  {journal} {\bibinfo  {journal} {Nat. Phys.}\ }\textbf {\bibinfo
  {volume} {8}},\ \bibinfo {pages} {887} (\bibinfo {year} {2012})}\BibitemShut
  {NoStop}%
\bibitem [{\citenamefont {Churchill}\ \emph {et~al.}(2013)\citenamefont
  {Churchill}, \citenamefont {Fatemi}, \citenamefont {Grove-Rasmussen},
  \citenamefont {Deng}, \citenamefont {Caroff}, \citenamefont {Xu},\ and\
  \citenamefont {Marcus}}]{Churchill2013}%
  \BibitemOpen
  \bibfield  {author} {\bibinfo {author} {\bibfnamefont {H.~O.~H.}\
  \bibnamefont {Churchill}}, \bibinfo {author} {\bibfnamefont {V.}~\bibnamefont
  {Fatemi}}, \bibinfo {author} {\bibfnamefont {K.}~\bibnamefont
  {Grove-Rasmussen}}, \bibinfo {author} {\bibfnamefont {M.~T.}\ \bibnamefont
  {Deng}}, \bibinfo {author} {\bibfnamefont {P.}~\bibnamefont {Caroff}},
  \bibinfo {author} {\bibfnamefont {H.~Q.}\ \bibnamefont {Xu}},\ and\ \bibinfo
  {author} {\bibfnamefont {C.~M.}\ \bibnamefont {Marcus}},\ }\href
  {https://doi.org/10.1103/PhysRevB.87.241401} {\bibfield  {journal} {\bibinfo
  {journal} {Phys. Rev. B}\ }\textbf {\bibinfo {volume} {87}},\ \bibinfo
  {pages} {241401} (\bibinfo {year} {2013})}\BibitemShut {NoStop}%
\bibitem [{\citenamefont {Albrecht}\ \emph {et~al.}(2016)\citenamefont
  {Albrecht}, \citenamefont {Higginbotham}, \citenamefont {Madsen},
  \citenamefont {Kuemmeth}, \citenamefont {Jespersen}, \citenamefont
  {Nyg{\aa}rd}, \citenamefont {Krogstrup},\ and\ \citenamefont
  {Marcus}}]{Albrecht2016}%
  \BibitemOpen
  \bibfield  {author} {\bibinfo {author} {\bibfnamefont {S.~M.}\ \bibnamefont
  {Albrecht}}, \bibinfo {author} {\bibfnamefont {A.~P.}\ \bibnamefont
  {Higginbotham}}, \bibinfo {author} {\bibfnamefont {M.}~\bibnamefont
  {Madsen}}, \bibinfo {author} {\bibfnamefont {F.}~\bibnamefont {Kuemmeth}},
  \bibinfo {author} {\bibfnamefont {T.~S.}\ \bibnamefont {Jespersen}}, \bibinfo
  {author} {\bibfnamefont {J.}~\bibnamefont {Nyg{\aa}rd}}, \bibinfo {author}
  {\bibfnamefont {P.}~\bibnamefont {Krogstrup}},\ and\ \bibinfo {author}
  {\bibfnamefont {C.~M.}\ \bibnamefont {Marcus}},\ }\href
  {https://doi.org/10.1038/nature17162} {\bibfield  {journal} {\bibinfo
  {journal} {Nature}\ }\textbf {\bibinfo {volume} {531}},\ \bibinfo {pages}
  {206} (\bibinfo {year} {2016})}\BibitemShut {NoStop}%
\bibitem [{\citenamefont {\"{O}nder G\"{u}l}\ \emph {et~al.}(2018)\citenamefont
  {\"{O}nder G\"{u}l}, \citenamefont {Zhang}, \citenamefont {Bommer},
  \citenamefont {de~Moor}, \citenamefont {Car}, \citenamefont {Plissard},
  \citenamefont {Bakkers}, \citenamefont {Geresdi}, \citenamefont {Watanabe},
  \citenamefont {Taniguchi},\ and\ \citenamefont {Kouwenhoven}}]{Guel2018}%
  \BibitemOpen
  \bibfield  {author} {\bibinfo {author} {\bibnamefont {\"{O}nder G\"{u}l}},
  \bibinfo {author} {\bibfnamefont {H.}~\bibnamefont {Zhang}}, \bibinfo
  {author} {\bibfnamefont {J.~D.~S.}\ \bibnamefont {Bommer}}, \bibinfo {author}
  {\bibfnamefont {M.~W.~A.}\ \bibnamefont {de~Moor}}, \bibinfo {author}
  {\bibfnamefont {D.}~\bibnamefont {Car}}, \bibinfo {author} {\bibfnamefont
  {S.~R.}\ \bibnamefont {Plissard}}, \bibinfo {author} {\bibfnamefont {E.~P.
  A.~M.}\ \bibnamefont {Bakkers}}, \bibinfo {author} {\bibfnamefont
  {A.}~\bibnamefont {Geresdi}}, \bibinfo {author} {\bibfnamefont
  {K.}~\bibnamefont {Watanabe}}, \bibinfo {author} {\bibfnamefont
  {T.}~\bibnamefont {Taniguchi}},\ and\ \bibinfo {author} {\bibfnamefont
  {L.~P.}\ \bibnamefont {Kouwenhoven}},\ }\href
  {https://doi.org/10.1038/s41565-017-0032-8} {\bibfield  {journal} {\bibinfo
  {journal} {Nat. Nanotechnol.}\ }\textbf {\bibinfo {volume} {13}},\ \bibinfo
  {pages} {192} (\bibinfo {year} {2018})}\BibitemShut {NoStop}%
\bibitem [{\citenamefont {Zhang}\ \emph {et~al.}(2018)\citenamefont {Zhang},
  \citenamefont {Liu}, \citenamefont {Gazibegovic}, \citenamefont {Xu},
  \citenamefont {Logan}, \citenamefont {Wang}, \citenamefont {van Loo},
  \citenamefont {Bommer}, \citenamefont {de~Moor}, \citenamefont {Car},
  \citenamefont {het Veld}, \citenamefont {van Veldhoven}, \citenamefont
  {Koelling}, \citenamefont {Verheijen}, \citenamefont {Pendharkar},
  \citenamefont {Pennachio}, \citenamefont {Shojaei}, \citenamefont {Lee},
  \citenamefont {Palmstr{\o}m}, \citenamefont {Bakkers}, \citenamefont
  {Sarma},\ and\ \citenamefont {Kouwenhoven}}]{Zhang2018}%
  \BibitemOpen
  \bibfield  {author} {\bibinfo {author} {\bibfnamefont {H.}~\bibnamefont
  {Zhang}}, \bibinfo {author} {\bibfnamefont {C.-X.}\ \bibnamefont {Liu}},
  \bibinfo {author} {\bibfnamefont {S.}~\bibnamefont {Gazibegovic}}, \bibinfo
  {author} {\bibfnamefont {D.}~\bibnamefont {Xu}}, \bibinfo {author}
  {\bibfnamefont {J.~A.}\ \bibnamefont {Logan}}, \bibinfo {author}
  {\bibfnamefont {G.}~\bibnamefont {Wang}}, \bibinfo {author} {\bibfnamefont
  {N.}~\bibnamefont {van Loo}}, \bibinfo {author} {\bibfnamefont {J.~D.~S.}\
  \bibnamefont {Bommer}}, \bibinfo {author} {\bibfnamefont {M.~W.~A.}\
  \bibnamefont {de~Moor}}, \bibinfo {author} {\bibfnamefont {D.}~\bibnamefont
  {Car}}, \bibinfo {author} {\bibfnamefont {R.~L. M.~O.}\ \bibnamefont {het
  Veld}}, \bibinfo {author} {\bibfnamefont {P.~J.}\ \bibnamefont {van
  Veldhoven}}, \bibinfo {author} {\bibfnamefont {S.}~\bibnamefont {Koelling}},
  \bibinfo {author} {\bibfnamefont {M.~A.}\ \bibnamefont {Verheijen}}, \bibinfo
  {author} {\bibfnamefont {M.}~\bibnamefont {Pendharkar}}, \bibinfo {author}
  {\bibfnamefont {D.~J.}\ \bibnamefont {Pennachio}}, \bibinfo {author}
  {\bibfnamefont {B.}~\bibnamefont {Shojaei}}, \bibinfo {author} {\bibfnamefont
  {J.~S.}\ \bibnamefont {Lee}}, \bibinfo {author} {\bibfnamefont {C.~J.}\
  \bibnamefont {Palmstr{\o}m}}, \bibinfo {author} {\bibfnamefont {E.~P. A.~M.}\
  \bibnamefont {Bakkers}}, \bibinfo {author} {\bibfnamefont {S.~D.}\
  \bibnamefont {Sarma}},\ and\ \bibinfo {author} {\bibfnamefont {L.~P.}\
  \bibnamefont {Kouwenhoven}},\ }\href {https://doi.org/10.1038/nature26142}
  {\bibfield  {journal} {\bibinfo  {journal} {Nature}\ }\textbf {\bibinfo
  {volume} {556}},\ \bibinfo {pages} {74} (\bibinfo {year} {2018})}\BibitemShut
  {NoStop}%
\bibitem [{\citenamefont {Sato}\ \emph {et~al.}(2009)\citenamefont {Sato},
  \citenamefont {Takahashi},\ and\ \citenamefont {Fujimoto}}]{Sato2009}%
  \BibitemOpen
  \bibfield  {author} {\bibinfo {author} {\bibfnamefont {M.}~\bibnamefont
  {Sato}}, \bibinfo {author} {\bibfnamefont {Y.}~\bibnamefont {Takahashi}},\
  and\ \bibinfo {author} {\bibfnamefont {S.}~\bibnamefont {Fujimoto}},\ }\href
  {https://doi.org/10.1103/PhysRevLett.103.020401} {\bibfield  {journal}
  {\bibinfo  {journal} {Phys. Rev. Lett.}\ }\textbf {\bibinfo {volume} {103}},\
  \bibinfo {pages} {020401} (\bibinfo {year} {2009})}\BibitemShut {NoStop}%
\bibitem [{\citenamefont {W\'ojcik}\ \emph {et~al.}(2018)\citenamefont
  {W\'ojcik}, \citenamefont {Bertoni},\ and\ \citenamefont
  {Goldoni}}]{Wojcik2018}%
  \BibitemOpen
  \bibfield  {author} {\bibinfo {author} {\bibfnamefont {P.}~\bibnamefont
  {W\'ojcik}}, \bibinfo {author} {\bibfnamefont {A.}~\bibnamefont {Bertoni}},\
  and\ \bibinfo {author} {\bibfnamefont {G.}~\bibnamefont {Goldoni}},\ }\href
  {https://doi.org/10.1103/PhysRevB.97.165401} {\bibfield  {journal} {\bibinfo
  {journal} {Phys. Rev. B}\ }\textbf {\bibinfo {volume} {97}},\ \bibinfo
  {pages} {165401} (\bibinfo {year} {2018})}\BibitemShut {NoStop}%
\bibitem [{\citenamefont {Groth}\ \emph {et~al.}(2014)\citenamefont {Groth},
  \citenamefont {Wimmer}, \citenamefont {Akhmerov},\ and\ \citenamefont
  {Waintal}}]{Groth2014}%
  \BibitemOpen
  \bibfield  {author} {\bibinfo {author} {\bibfnamefont {C.~W.}\ \bibnamefont
  {Groth}}, \bibinfo {author} {\bibfnamefont {M.}~\bibnamefont {Wimmer}},
  \bibinfo {author} {\bibfnamefont {A.~R.}\ \bibnamefont {Akhmerov}},\ and\
  \bibinfo {author} {\bibfnamefont {X.}~\bibnamefont {Waintal}},\ }\href
  {http://stacks.iop.org/1367-2630/16/i=6/a=063065} {\bibfield  {journal}
  {\bibinfo  {journal} {New J. Phys.}\ }\textbf {\bibinfo {volume} {16}},\
  \bibinfo {pages} {063065} (\bibinfo {year} {2014})}\BibitemShut {NoStop}%
\end{thebibliography}%
\end{document}